\documentstyle[12pt]{article}
\def\vsig{\mbox{\boldmath{$\sigma$}}}
\begin{document}
\begin{center}
An Extension of Fractional Parentage Expansion to the\\
Nonrelativistic and Relativistic $SU^f (3)$ Dibaryon Calculations\\

\vspace*{0.25in}

Fan Wang$^{a)}$, Jia-lun Ping$^{b)}$, and T.Goldman$^{c)}$\\

\end{center}
\vspace*{0.10in}

\noindent $^{a)}$ Department of Physics and Center for Theoretical
Physics, Nanjing University\\
\indent Nanjing, 210008, China\\
\noindent $^{b)}$ Department of Physics, Nanjing Normal University,
Nanjing, 210024, China\\
\noindent $^{c)}$ Theoretical Division, LANL, Los Alamos, NM87545, USA

\vspace*{-0.20in}

\begin{abstract}
The fractional parentage expansion method is extended from $SU^f(2)$
nonrelativistic to $SU^f(3)$ and relativistic dibaryon calculations.  A
transformation table between physical bases and symmetry bases for the
$SU^f(3)$ dibaryon is provided.  A program package is written for
dibaryon calculation based on the fractional parentage expansion method.
\end{abstract}

\pagebreak

\begin{center}
\large{{\bf I. INTRODUCTION}}
\end{center}

Quantum chromodynamics (QCD) is a very promising theory for
fundamental strong interaction.  However, due to the complexity of QCD,
for the present time and for the foreseeable future, one must rely on QCD
inspired models to study hadron physics.  The existing models (potential,
bag, soliton, etc.) are quite successful for the meson and baryon sectors,
but not so successful for hadronic interactions.  Recently some hope has
developed to obtain the full N-N interaction from QCD models[1,2].

Since the first prediction of H particle by Jaffe[3], there have been
tremendous efforts both theoretically and experimentally[4] to find possible
candidates for quasi-stable dibaryon states.  Nevertheless, there remains
an outstanding question.  Theoretically all the QCD models, including
lattice QCD calculations, predict that there should be quasi-stable
dibaryons or dibaryon resonances, but in contrast, experimentally no
quasi-stable dibaryon whatsoever has been observed (except the
molecular deuteron state).  One has to ask if some important QCD
characteristics are missing in all these dibaryon calculations.  For example
in the potential (or cluster) model approach, the six quark Hamiltonian is
usually a direct extension of the three quark Hamiltonian.  This extension
is neither reasonable nor successful.  The two-body confinement potential
yields color van der Waals forces which are in contradiction with
experimental observation.  Lattice gauge calculations and nonperturbative
QCD both yield a string-like structure inside a hadron instead of two-body
confinement.  Two-body confinement may be a reasonable approximation
inside a hadron but not for the interaction between quarks in two color
singlet hadrons[5].  Another possible missed general feature is that the
quark, originally confined in a single hadron, may tunnel (or percolate) to
the other hadron when two hadrons are close together[6].  In the potential
model approach, the internal motion of the interacting hadrons is assumed
to be unchanged.  The product ansatz of the Skyrmion model approach
makes the same approximation.  In the bag model approach, another
extreme approximation is assumed, {\it i.e.}, the six quark are merged into
a single confinement space. The real configuration may be in between
these two extremes which is well known in the molecular physics.

Except for a few cluster model calculations, in which a phenomenological
meson exchange ls involved to fit the N-N scattering, for all the other
dibaryon calculations, the model parameters are only constrained by
hadron spectroscopy.  In fact, the six quark system includes new color
structures, for which a single hadron cannot give any information.  A six
quark Hamiltonian should be  constrained by the existing baryon-baryon
interaction data, especially the N-N data, then the model dibaryon states
may be really relevant to the experimental measurement.

A model, the quark delocalization color screening model(QDCSM), has
been developed which includes the new QCD inspired ingredients
mentioned above and is constrained by N-N scattering data[2].  This
model has been applied to a systematic search of the dibaryon candidates
in the $u$ $d$ and $s$ three flavor world[7] to provide a better estimate of
dibaryon states on the one hand and to test the model assumption further
on the other hand.

As pointed out in [4], a more realistic systematic search of dibaryons
would be a tremendous task, for which a systematic and powerful method
is indispensable.  The fractional parentage (fp) expansion developed in
atomic and nuclear physics is one of such methods.  A major obstacle in
applying the fp expansion technique to quark models is the occurrence of
many $SU(mn) \supset SU(m) \times SU(n)$ isoscalar factors(ISF) with
$m, n \geq 2$. e.g., in the two ``orbits'', two spins, $n_f$ flavors and 3
colors quark world.  We need the $SU(2 \times 3 \times n_f \times 2)
\supset SU^x (2) \times SU(6n_f), SU(6n_f) \supset SU^c (3) \times
SU(2n_f ), SU(2n_f) \supset SU^f (n_f ) \times SU^{\sigma} (2)$ ISF 's,
where $x, c, f$ and $\sigma$ indicate the space or orbit, color, flavor and
spin respectively.  Before 1991, only the $SU(4) \supset SU(2) \times
SU(2)$ ISF (and some scattered results for $SU(6) \supset SU(3) \times
SU(2)$ ISF were available.  A breakthrough in group representation
theory is the recognition of the fact that the $SU(n_1 n_2) \supset SU(n_1)
\times SU(n_2)$
$n_2$-particle coefficients of $fp$($cfp$) are precisely the ISF for the
permutation group chain $S(n_1 + n_2) \supset S(n_1) \times S(n_2)$[8a],
and the former can be calculated and tabulated in a rank independent
way, instead one $m$ and $n$ at a time.  In 1991, Chen {\it et al}[8b]
published a book with phase consistent $SU(mn) \supset SU(m) \times
SU(n)$ ISF for arbitrary $m$ and $n$ and for up to six particles.  Because
of this, we are now in a position to develop an efficient algorithm for
dibaryon calculations based on the $fp$ technique.  This paper reports the
extension of the $fp$ expansion to the nonrelativistic and relativistic $SU^f
(3)$ quark model calculation in line with the work of Harvey[9] and of
Chen[l0].

\begin{center}

\large{{\bf II. PHYSICAL BASES AND SYMMETRY BASES}}
\end{center}

A dibaryon may be a loosely bound two $q^3$ cluster state like the
deuteron, or it may be a tightly bound $q^6$ cluster like the Jaffe's version
of the H particle.  Many cases may be in between.  To describe these
states, the physical basis is preferable due to its apparent dibaryon
content in
the asymptotic region without artificial confinement assumptions.  The
physical basis is nothing else
but the cluster model basis developed in the nuclear cluster model[11]. To
show the symmetry property explicitly, we follow Chen's notation[8,10] but
with a slight modification, because we are working in the $u, d, s$ three
flavor world instead of $u, d$ flavors.

A baryon in the $u, d, s$ three flavor world is described by \\

$$
\psi (B) = \left| \stackrel{[\nu]}{[\sigma]W[\mu][f]YIJ}\right>, \eqno(1)
$$
which is a basis vector belonging to the irreducible representations (irreps)
\begin{eqnarray*}
\stackrel{[1^3]}{SU(36) \supset} \; \left(\stackrel{[\nu]}{SU^x(2)} \times
\left(\stackrel{[\tilde{\nu}]}{SU(18)}\supset \stackrel{[\sigma]}{SU^c(3)}
\times \left(\stackrel{[\mu]}{SU(6)} \supset \right. \right. \right.\\
\left. \left. \left. \left(\stackrel{[f]}{SU^f(3) }\supset \;
\stackrel{I}{SU^{\tau}(2)} \times  \stackrel{Y}{U^Y(1)}\right) \times
\stackrel{J}{SU^{\sigma}(2)}\right) \right) \right),
\end{eqnarray*}
\vspace*{-0.50in}\\
\hspace*{5.70in} (2)\\

\normalsize
\noindent where the first reduction is to orbital times combined
color-flavor-spin symmetry, the second reduces the latter to color times
combined flavor-spin, and the third reduces the last to flavor (which is
itself reduced to isospin times hypercharge) times spin.  Here $[\nu]$ etc.\
are the Young diagrams describing the permutational and $SU(n)$
symmetries.  In our calculation, the ground state baryons are assumed to
be in the totally symmetric orbital state $[\nu] = [3]$, while $[\sigma] =
[1^3] W$ is the Weyl tableau for the $SU^c(3)$ state due to color
confinement, i.e.,
the baryon is colorless.  On the other hand, $[\mu] = [3]$ due to the totally
antisymmetry requirement $[\sigma] \times [\mu] \rightarrow [\tilde{\nu}] =
[1^3], [\tilde{\nu}]$ being the conjugate Young diagram of $[\nu]$.  $[f]$
and $[\sigma_J]$ are restricted by the condition $[f] \times [\sigma_J]
\rightarrow [\mu] = [3]$, this leads to $[f] = [\sigma_J] = [3]$ or $[21],
[\sigma_J]$ represents the spin symmetry, $[\sigma_J] = [\frac{n}{2} + J,
\frac{n}{2} - J], n$ is the total number of quarks. i.e., the $SU^f(3)$
decuplet and octet baryons $\Delta, \Sigma^*, \Xi^*, \Omega \; {\rm and} \;
N, \Lambda, \Sigma, \Xi$. The symbols $Y, I, J$ denote the hypercharge,
isospin and spin quantum numbers respectively.  $M_I$ and $M_J$, the
magnetic quantum numbers, are omitted in eq.(1).

A two baryon physical basis is described by

\begin{eqnarray}
\Psi_{\alpha k} (B_1 B_2) &=& {\cal A} [\psi (B_1) \psi (B_2)]
\stackrel{[\sigma]}{W} \stackrel{\;\;\;I\;\;\;\;\;J\;\;\;}{M_I M_J} \nonumber
\\
&=& {\cal A} \left[ \left.\left| \;
\stackrel{[\nu_1]}{[\sigma_1][\mu_1][f_1]Y_1I_1J_1} \right> \right| \left.
\stackrel{[\nu_2]}{[\sigma_2][\mu_2][f_2]Y_2I_2J_2} \right> \right] \;
^{[\sigma]\;\;I\;\;\;J\;\;\;}_{WM_IM_J}, \hspace*{.750in} (3) \nonumber
\end{eqnarray}
here $[\;]\;^{[\sigma]\;I\;\;\,\;J\;\;}_{W M_I M_J}$means the couplings in
terms of the $SU^c(3), SU^{\tau}(2)$ and $SU^{\sigma}(2)$
Clebsch-Gordan coefficients (CGC) so that it has the total color symmetry
$[\sigma]W$, isospin $I M_I$ and spin $JM_J$.
Due to color confinement, only the overall color singlet $[\sigma] = [2^3]$
is allowed.  ${\cal A \rm}$ is a normalized antisymmetric operator.  $\alpha
= (YIJ)$ with $Y = Y_1 + Y_2, k$ represents the quantum numbers $\nu_i,
\sigma_i, \mu_i, f_i, I_i, J_i (i = 1,2)$.

To take into account the mutual distortion or the internal orbital excitation
of the interacting baryons when they are near one another, the delocalized
single quark state $l(r)$ is used for baryon $B_1 (B_2) [2,6,7]$
$$
l = (\phi_L + \epsilon (s) \phi_R) \left/ N (s), r = (\phi_R + \epsilon (s)
\phi_L) \right/ N(s)
$$
$$
\phi_L (\vec{r}) = \left( \frac{1}{\pi b^2}\right)^{\frac{3}{4}} e^{-
\frac{(\vec{r}
- \vec{s}/2)^2}{2b^2}}, \eqno (4)
$$
$$
N^2(s) = 1 + \epsilon^2 (s) + 2 \epsilon (s) \left< \phi_L | \phi_R \right> .
$$
The $\vec{s}$ is the separation between two $q^3$ cluster centers,
$\epsilon (s)$ is a parameter describing the delocalization (or percolation)
effect, and it is determined variationally by the $q^6$ dynamics.  Hidden
color channels are not included in eq.(3), because it has been proven[12]
that the colorless hadron channels form a complete Hilbert space if the
excited colorless baryon states are included.  Also the concept of a
colorful hadron has not been well defined in QCD models.

Physical bases are not convenient for matrix element calculations.  To
take advantage of the $fp$ expansion technique developed in atomic and
nuclear physics, one has to use symmetry bases (group chain
classification bases). This requires an extension of the $q^3$ state eq.(l)
to the $q^6$ case,
$$
\Phi_{\alpha K} (q^6) = \left |
\stackrel{[\nu]l^3r^3}{[\sigma]W[\mu]\beta[f]YIJM_IM_J}\right>. \eqno(5)
$$
Here $K$ represents the quantum numbers $[\nu], [\mu], \beta, [f]$
appeared in eq.(5).  $[\sigma] = [2^3]$ due to color confinement. To be
consistent with the physical basis choice, the orbital part is truncated to
include the $l^3r^3$ configuration only.  $[\nu]$ is restricted to be $[\nu] =
[3] \times [3] = [6] + [51] + [42] + [33]$.  $\beta$ is the inner multiplicity
index in the reduction $[\mu] \rightarrow [f] \times [\sigma_J]$.

Physical and symmetry bases both form a complete set in a truncated
Hilbert space, and are related by a unitary transformation.  Harvey[9] first
calculated the transformation coefficients for the $u, d$ two flavor case.
Chen[l0] proved that the transformation coefficients are just a product of
$(6 \rightarrow 3 + 3) SU(mn) \supset SU(m) \times SU(n)$ isoscalar
factors.  Here we extend them to the $SU^f(3)$ case,
$$
{\cal A \rm} \left[ \left| \left. \left.
\stackrel{[\nu_1]l^3}{[\sigma_1][\mu_1][f_1]
Y_1I_1J_1} \right> \right| \stackrel{[\nu_2]r^3}{[\sigma_2][\mu_2][f_2]
Y_2I_2J_2} \right> \right]\; ^{[\sigma]\;\;I\;\;\;J\;\;\;}_{WM_IM_J} \eqno(6)
$$
$$
= \sum_{\tilde \nu \mu \beta f \gamma} \; C \;
^{[\bar{\nu}][\sigma][\mu]}_{[\bar{\nu}_1][\sigma_1][\mu_1][\bar{\nu}_2]
[\sigma_2][\mu_2]} \; C \; ^{[\mu]\beta[f]\gamma J}_{[\mu_1][f_1]
J_1,[\mu_2][f_2] J_2} \; C \; ^{[f]\gamma Y I}_{[f_1] Y_1 I_1 [f_2] Y_2
I_2} \left| \left. ^{[\nu] l^3 r^3}_{[\sigma] W [\mu] \beta [f] Y I J
M_I M_J} \right.  \right> .
$$
\normalsize
This expression is written simply as
$$
\Psi_{\alpha k} (B_1B_2) = \sum_K C_{k K} \Phi_{\alpha K} (q^6), \eqno(7)
$$
here $\gamma$ is an outer multiplicity index in the reduction $[f_1]
\times [f_2]  \rightarrow [f]$.   The first two C factors in Eq.(6) are
$SU(18)\supset SU^c(3) \times  SU(6)$ and $SU(6) \supset SU^f (3)\times
SU^{\sigma}(2)$ isoscalar factors respectively and the third one is
$SU^f (3) \supset SU^{\tau}(2) \times U^Y(1)$ isoscalar factor.  All
these isoscalar factors can be found in Chen's book[8b]. The calculated
transformation coefficients are listed in table 1.  The $Y = 2$ part is
a revised version (phase consistent and simplified for $I = J = 1$
case) of Harvey's table 11[9].  (The relationship between our Tables
and those of Harvey, is discussed in the Appendix.)  The $Y \neq 2$
part is an extension of Harvey's two flavor case to three flavor case.
Because the hidden color channels are not included, this table can be
used to expand the physical bases in terms of the symmetry bases only.
If one wants to expand the symmetry bases in terms of the physical
bases, then the hidden color physical bases (or other equivalent set of
bases) should be added.  One example is given below,

\begin{eqnarray*}
\hspace*{.4in} \left. | \; H \; \right >  & = &  \sqrt{\frac{1}{5}}
\left ( \sqrt{\frac{3}{8}} \left. | \Sigma \Sigma  \right > -
\sqrt{\frac{4}{8}} \left. | \bar{N \Xi} \right > - \sqrt{\frac{1}{8}} |
\Lambda \Lambda > \right ) \nonumber \\ & - & \sqrt{\frac{6}{40}}
\left. | \Sigma \Sigma \right >_c + \sqrt{\frac{8}{40}} \left. | \bar{N
\Xi} \right >_c + \sqrt{\frac{2}{40}} \left. | \Lambda \Lambda \right
>_c \nonumber \\ & + & \sqrt{\frac{3}{40}} \left. | \Sigma^{\prime}
\Sigma^{\prime} \right >_c - \sqrt{\frac{4}{40}} \left. |
\bar{N^{\prime} \Xi^{\prime}} \right >_c - \sqrt{\frac{1}{40}} \left. |
\Lambda^{\prime} \Lambda^{\prime} \right >_c - \sqrt{\frac{8}{40}}
\left. | \Lambda_s \Lambda_s \right >_c. \hspace*{.4in}  (8)
\end{eqnarray*}
Here $|\overline{XY}\rbrace$ means the symmetric channel of baryons $X$
and $Y$, $|XY \rangle_c$ means hidden color channel of colorful baryons
$X$ and $Y$, $\Lambda_s$ is the flavor singlet $\Lambda$, $X^{\prime}$
represents excited colorful baryon with spin $\frac{3}{2}$. In the
prevailing literature, only first three colorless channels are
given[13].  See the Appendix for a description of the difference
between our meaning for symmetry and that of Harvey [9].

\begin{center}
\large{{\bf III FRACTIONAL PARENTAGE EXPANSION}}
\end{center}

A physical six quark state with quantum number $\alpha = (YIJ)$ is
expressed as a channel coupling wave function (WF)
$$
\Psi_{\alpha} = \sum_{k} C_k \Psi_{\alpha k} (B_1 B_2). \eqno (9)
$$
The channel coupling coefficients $C_k$ are determined by the
diagonalization of the six quark Hamiltonian as usual.
To calculate the six quark Hamiltonian matrix elements in the physical
basis,
$$
H_{kk^{\prime}} = \left \langle \Psi_{\alpha k} || H || \Psi_{\alpha
k^{\prime}} \right \rangle, \eqno (10)
$$
is tedious. We first express the physical basis in terms of the
symmetry basis by the transformation eq.(6), and the matrix element
eq.(l0) is transformed into a sum of matrix elements in the symmetry
basis
$$
H_{k k^{\prime}} = \sum_{K, K^{\prime}} C_{kK} C_{k^{\prime}K^{\prime}}
\left \langle \Phi_{\alpha K} |H| \Phi_{\alpha K^{\prime}} \right
\rangle.  \eqno (11)
$$
The matrix elements $\left \lbrace \Phi_{\alpha K} |H| \Phi_{\alpha
K^{\prime}} \right \rbrace$ can be calculated by the well known fp
expansion method,
$$
\left \langle \Phi_{\alpha K} |H| \Phi_{\alpha K^{\prime}} \right
\rangle = \sum \left(^6_2 \right) \left \langle \Phi_{\alpha K} |
\alpha_1 K_1, \alpha_2 K_2 \right \rangle \left \langle
\alpha^{\prime}_1 K^{\prime}_1, \alpha^{\prime}_2 K^{\prime}_2 |
\Phi_{\alpha K^{\prime}} \right \rangle \left \langle \alpha_1 K_1 |
\alpha^{\prime}_1 K^{\prime}_1 \right \rangle
$$
$$
\left \langle \alpha_2 K_2 | H_{56} | \alpha^{\prime}_2 K^{\prime}_2
\right \rangle .  \eqno (12)
$$
Here $\left \langle \alpha_1 K_1 | \alpha^{\prime}_1 K^{\prime}_1
\right \rangle$ is the four quark overlap, and is a little more
complicated than the atomic and nuclear shell model case due to the
non-orthogonal property of the single quark orbital state (see below).
$\left \langle \alpha_2 K_2 | H_{56} | \alpha^{\prime}_2 K^{\prime}_2
\right \rangle$ is the two body matrix element and $H_{56}$ represents
the two-body operator for the last pair.  $\left( ^6_2 \right) = 15$ is
the interacting pair number.  To simplify the computer program, the
one-body operator matrix elements are calculated by the same expansion
eq.(12) with the obvious substitution $H_{56} \rightarrow H_5 + H_6$
and $\left(^6_2 \right) \rightarrow \frac{6}{2} = 3$ (only six one-body
operators altogether instead of 15 pair interactions).

$\left \langle \Psi_{\alpha K} | \alpha_1 K_1, \alpha_2 K_2 \right
\rangle$ and $\left \langle \alpha^{\prime}_1 K^{\prime}_1,
\alpha^{\prime}_2 K^{\prime}_2 | \Phi_{\alpha K^{\prime}}\right
\rangle$ are the total Clebsch-Gordon Coefficients (CGC). They are
calculated as follows[8b].
$$
\left \langle \hspace*{-0.025in}
\stackrel{[\nu]l^3r^3}{[\sigma]W[\mu]\beta[f]YIJM_IM_J}
\hspace*{-0.05in} \left|
\stackrel{[\nu_1]W^x_1}{[\sigma_1]W^c_1,[\mu_1][f_1]Y_1I_1J_1M_{I_{1}}
M_{J_{1}}}, \stackrel{[\nu_2]W^x_2}{[\sigma_2]
W^c_2[\mu_2][f_2]Y_2I_2J_2M_{I_{2}}M_{J_{2}}} \hspace*{-0.025in} \right
\rangle \right.
$$
$$
= \sum_{\gamma} \; C \; ^{[\sigma] W}_{[\sigma_1] W^c_1\, , \,
[\sigma_2] W^c_2} \; C \; ^{IM_I}_{I_1 M_{I_{1}}\, , \,I_2 M_{I_{2}}}
\; C \; ^{JM_J}_{J_1 M_{J_{1}} \, , \, J_2 M_{J_{2}}} \; C \;
^{[\nu]l^3 r^3}_{[\nu_1]{W^x_1} \, , \, [\nu_2] W^x_2}
$$
$$
\hspace*{0.350in}C \; ^{[1^6][\nu][\tilde{\nu}]\hspace*{0.64in}}_{[1^
4][\nu_1][\tilde{\nu}_1] , [1^2][\nu_2][\tilde{\nu}_2]} \; C \;
^{[\tilde{\nu}][\sigma][\mu]\hspace*{0.64in}}_{[\tilde{\nu}_1][\sigma_1][\mu_
1] , [\tilde{\nu}_2][\sigma_2][\mu_2]} \; C \; ^{[\mu]\beta[f]\gamma J
\hspace*{0.64in}}_{[\mu_1][f_1][J_1] , [\mu_2][f_2][J_2]} \; C \;
^{[f]\gamma YI \hspace*{0.64in}}_{[f_1]Y_1 I_1,[f_2] Y_2 I_2}.
$$
\normalsize
\hspace*{5.70in}(13)

\noindent The first four C's are the $SU^c (3), SU^{\tau} (2)$,
$SU^{\sigma}(2)$ and $SU^x (2)$ CGC, the next three C's are the $SU(36)
\supset SU^x (2) \times SU(18), SU( 18)\supset SU^c (3) \times SU(6),
SU(6) \supset SU^f (3) \times SU^{\sigma}(2)$ isoscalar factors, the
last one is the $SU^f (3) \supset SU^{\tau}(2) \times U^Y (1)$
isoscalar factor. All these isoscalar factors (for particle number
$\leq$ 6) can be found in ref.[8b].  The $SU^x$(2) orbital CGC
$C^{[\nu]l^3r^3}_{[\nu_1]W^x_1 , [\nu_2] W^x_x}$ is called the orbital
two-body cfp by Harvey and listed in his table 4[9].  It is obvious
that it is better to use the standard phase convention of the SU(2)
CGC.  Then the entries under [4] : [ll] $a^2 b^2 : \tilde{ab}$, [31] :
[2] $ab^3 : a^2$ and [31] : [2]$a^2b^2 : \bar{ab}$ should be assigned
opposite signs.

The four-quark state $\left| \left. \alpha_1 K_1 \right \rangle
\right.$ can be expressed as[l0]
$$
\left| \left. \alpha_1K_1\right \rangle \right. = \left| \left.
\stackrel{[\nu_1]W^x_1}{[\sigma_1]W^c_1[\mu_1][f_1]Y_1I_1J_1M_{I_{1}}
M_{J_{1}}} \right. \right \rangle \eqno (14)
$$
$$
= \left. \left. \sum_m (h_{\nu_{1}})^{-\frac{1}{2}}\Lambda^{\nu_1}_m
\left| \;^{[\nu_1]W^x_1}_{m} \; \right \rangle \right| \;
^{[\tilde{\nu}_1]}_{\tilde{m}} \; [\sigma_1]W^c_1
[\mu_1][f_1]Y_1I_1M_{I_{1}} M_{J_{1}} \right \rangle,
$$
here $m(\tilde{m})$ is the Yamanouchi number of the Young tableau,
$(h_{\nu_{1}})^{-\frac{1}{2}}\Lambda^{\nu_{1}}_m$ is the CGC for
$[\nu_1] \times [\tilde{\nu}_1] \rightarrow [1^4]$ of the permutation
group.  The color-flavor-spin part
$$
\left| \left. ^{[\tilde{\nu}_1]}_{\tilde{m}} [\sigma_1] W^c_1 [\mu_1]
[f_1] Y_1 I_1 J_1 M_{I_{1}} M_{J_{1}}  \right. \rangle \right.
$$
\noindent is orthogonal as usual
$$
\left \langle ^{[\tilde{\nu}_1]}_{\tilde{m}} [\sigma_1] W^c_1 [\mu_1]
[f_1] Y_1 I_1 J_1 M_{I_{1}} M_{J_{1}} \left|
^{[\tilde{\nu}^{\prime}_1]}_{\tilde{m}^{\prime}} [\sigma^{\prime}_1]
W^{\prime c}_1 [\mu^{\prime}_1] [f^{\prime}_1] Y^{\prime}_1
I^{\prime}_1 J^{\prime}_1 M^{\prime}_{I_{1}} M^{\prime}_{J_{1}} \right
\rangle = \right.  \delta_{1 1^{\prime}} \eqno (15)
$$
here $\delta_{1 1^{\prime}}$ is a product of the $\delta_{\nu_{1}
\nu^{\prime}_1}, \delta_{m m^{\prime}}$, $\cdot \cdot \cdot$, which
includes every pair of quantum numbers.  The only complication is
caused by the non-orthogonality of the single quark orbital state,
$$
\left \langle ^{[\nu_1]W^x_1}_{m} \bigg{|} \; ^{[\nu^{\prime}_1]
W^{\prime x}_{1}}_{m^{\prime}}  \right \rangle  =
\delta_{\nu_{1}\nu^{\prime}_{1}} \delta_{m m^{\prime}} \left \langle
^{[\nu_1]W^x_1}_{m} \bigg{|} \; ^{[\nu^{\prime}_1] W^{\prime
x}_{1}}_{m^{\prime}} \right \rangle \eqno (16)
$$
Finally we have the four quark overlap
\begin{eqnarray*}
\hspace*{1.20in} \left \langle
\alpha_1K_1|\alpha^{\prime}_1K^{\prime}_1 \right \rangle &=&
\delta_{11^{\prime}} h^{-1}_{\nu_{1}} \sum_m \left \langle
^{[\nu_1]W^x_1}_{m} \bigg{|} \; ^{[\nu_1]W^{\prime x}_1}_{m} \right
\rangle \\ \nonumber &=& \delta_{1 1^{\prime}} \left \langle
^{[\nu_1]W^x_1}_{m} \bigg{|} \; ^{[\nu_1] W^{\prime x}_{1}}_{m} \right
\rangle (\rm{any \; m}) \hspace*{1.20in} (17)
\end{eqnarray*}
This four-body overlap is listed by Harvey in his table 6[9].  To be
consistent with the standard $SU^x$(2) CGC phase convention, all the
entries in his table 6 should have positive signs.  Another
modification is caused by the delocalized orbit eq.(4):   The $m$ in
Harvey's table 6 should be replaced by
$$
m \rightarrow (2 \epsilon + (1 + \epsilon^2)F)/(1 + \epsilon^2 +
2\epsilon F), F = \left \langle \phi_L | \phi_R \right \rangle.
$$
Harvey's result is our $\epsilon = 0$ limit.

The two-quark state
$$
\hspace*{1.30in}| \left. \alpha_2K_2 \; \right \rangle \left. = \left|
\stackrel{[\nu_2]W^x_2}{[\sigma_2]W^c_2[\mu_2][f_2]Y_2I_2J_2M_{I_{2}}
M_{J_{2}}} \right. \right \rangle \hspace*{1.30in} (18)
$$
can be expressed in a similar form as eq.(14).  But the $[\nu_2],
[\sigma_2],[\mu_2], \rm{and} [f_2]$ are either symmetric [2] or
antisymmetric[$1^2$], and eq.(18) is in fact just a product of orbital,
color, flavor, and spin part.  The two-body interaction matrix elements
can be factorized too,
$$
\hspace*{-0.20in} \langle
\alpha_2K_2|H_{56}|\alpha^{\prime}_2K^{\prime}_2 \rangle =
\left \langle ^{[\nu_2]}_{W^x_2} \bigg{|} H^x_{56} \bigg{|} \;
^{[\nu^{\prime}_2]}_{W^{\prime x}_2} \; \right \rangle \; \left \langle
^{[\sigma_2]}_{W^c_2} \bigg{|} H^c_{56} \bigg{|} \;
^{[\sigma^{\prime}_2}_{W^{\prime x}_{W^{\prime c}_{2}}} \right \rangle
$$
$$
\hspace*{2.3950in} \left \langle ^{[f_2]}_{Y_{2}I_{2}M_{I_{2}}}
\bigg{|} H^f_{56} \bigg{|} \; ^{[f^{\prime}_2]}_{Y^{\prime}_2
I^{\prime}_2 M^{\prime}_{I_{2}}} \; \right \rangle \langle J_2
M_{J_{2}} | H^{\sigma}_{56} | J^{\prime}_2 M^{\prime}_{J_{2}} \rangle .
\hspace*{0.30in}(19)
$$

Here we have used the fact that the two body interaction is a sum of
terms of the form which we take as a single term for simplicity below.
$$
\hspace*{1.3950in}H_{56} = H^c_{56} \cdot H^f_{56} \cdot
H^{\sigma}_{56} \cdot H^x_{56} \hspace*{2.50in} (20)
$$

For the nonrelativistic case, $H$ is a scalar of $SU^c(3),
SU^{\tau}(2)$ and $SU^{\sigma}(2)$, the two-body matrix elements are
$W^c_2, M_{I_{2}}$ and $M_{J_{2}}$ independent, and the first three CGC
in eq.(13) will disappear in the matrix element $\langle \Phi_{\alpha
K} | H | \Phi_{\alpha K^{\prime}} \rangle$ of eq.(12) due to the
orthonormal property of CGC's.

For the one body operator (kinetic energy in a nonrelativistic model,
kinetic energy and mean field in a relativistic model)
$$
H_{56} = H_5 + H_6 ,
$$
by expanding the coupled state into the product of two particle states
with CGC and using the orthonormal property of CGC, the two one-body
operator matrix elements can be calculated very easily.  The $6
\rightarrow 5 + 1 fp$ expansion can be avoided and only the $6
\rightarrow 4 + 2 fp$ coefficients need to be included in a computer
program package.

\begin{center}
\large{{\bf IV THE RELATIVISTIC EXTENSION}}
\end{center}

It is commonly believed that the classification scheme eq.(2) can be
applied to the nonrelativistic quark only, because the spin and orbital
part are intrinsically coupled into a Dirac spinor for a relativistic
quark.  However in a Dirac cluster model, only the lowest Dirac state
is used and the lowest state of a Dirac particle moving in a central
potential can be expressed as a product of a pseudo-orbit and a Pauli
spinor[14]
$$
\phi_{\sigma}({\bf r}^{\prime}) = \left(^{\hspace*{0.20in} \phi_u({\bf
r}^{\prime})}_{-i \, \vsig  \, \cdot \hat{{\bf r}}^{\prime}
\phi_{\alpha^{\prime}}({\bf r}^{\prime})} \right)
\chi_{\sigma}Y_{00}(\theta^{\prime}, \phi^{\prime}). \eqno (21)
$$
Here ${\bf r}^{\prime} = {\bf r}-{\bf s}/2$ or ${\bf r} + {\bf s}/2$
depends on the confinement center, $\chi_{\sigma}$ is the usual Pauli
spinor, $\sigma = j_z = \pm 1/2, Y_{00} = \sqrt{\frac{1}{4 \pi}},
\phi_u$ and $\phi_d$ are the upper and lower (down) components of the
Dirac WF.  Taking the $\sqrt{\frac{1}{4 \pi}} \left(
^{\hspace*{0.20in}\phi_u ({\bf r}^{\prime})}_{- i \vsig \cdot \hat{{\bf
r}}^{\prime} \phi_d ({\bf r}^{\prime})} \right)$ as a pseudo-orbit WF
equivalent to that for the nonrelativistic orbital WF, we obtain two
linear independent states as the bases of a pseudo-orbit $SU^x (2)$ for
the Dirac quark.  In this way we can use the same classification scheme
eq.(2) to describe the six Dirac quark system[15].  The whole
calculation method discussed in Sec.\ II and III can be extended to a
Dirac quark cluster model directly.  The only difference is that when
we calculate the one- and two-body matrix elements, we have to
recombine the pseudo-orbit and the Pauli spinor together to be a Dirac
spinor.  For the four quark overlap calculation, recombination of the
pseudo-orbit and Pauli spinor seems to be needed too.  However, because
we only use the lowest Dirac state WF eq.(21), the single particle
overlap still can be separated into a pseudo-orbit part and a Pauli
spinor part
\begin{eqnarray*}
< \psi_{\sigma_{1}} ({\bf r}_1) | \psi_{\sigma_{2}} ({\bf r}_2) &=&
\chi^{\dagger}_{\sigma_{1}} \frac{1}{4 \pi} \int d{\bf r} (\psi_u ({\bf
r}_1), i \vsig \cdot {\bf r}_1 \psi_d ({\bf r}_1)) (\psi_u ({\bf r}_2),
- i \vsig \cdot {\bf r}_2 \psi_d ({\bf r}_2)) \chi_{\sigma_{2}}
\\ \nonumber &=& \chi^{\dagger}_{\sigma_{1}} \frac{1}{4 \pi} \int d{\bf
r} [\psi_u ({\bf r}_1) \psi_u ({\bf r}_2) + \vsig \cdot {\bf r}_1 \vsig
\cdot {\bf r}_2 \psi_d ({\bf r}_1) \psi_d ({\bf r}_2)]
\chi_{\sigma_{2}} \\ \nonumber &=& \chi^{\dagger}_{\sigma_{1}}
\chi_{\sigma_{2}} \frac{1}{4 \pi} \int d{\bf r} [\psi_u ({\bf r}_1)
\psi_u ({\bf r}_2) + \hat{{\bf r}}_1 \cdot \hat{{\bf r}}_2 \psi_d ({\bf
r}_1) \psi_d ({\bf r}_2)]. \hspace*{0.50in} (22)
\end{eqnarray*}
The spin dependent part is identically zero[16],
$$
\frac{1}{4 \pi} \int i \vsig \cdot (\hat{{\bf r}}_1 \times \hat{{\bf
r}}_2) \psi_d ({\bf r}_1) \psi_d ({\bf r}_2) d{\bf r} \equiv 0. \eqno
(23)
$$
Therefore the four quark overlap calculation can be done in an exactly
same way as that for the nonrelativistic case, i.e., separated into a
pseudo-orbital part and a spin part.

\begin{center}
\large{\bf V.  COMPUTERIZED FRACTIONAL PARENTAGE EXPANSION}
\end{center}

All the needed $SU(mn) \supset SU(m) \times SU(n)$ isoscalar factors
can be found directly from Chen's book[8b], the needed $SU(3) \supset
SU(2) \times U(1)$ isoscalar factors can be obtained from Chen's
$SU(3)$ CGC[8b] and the standard $SU(2)$ CGC.  (Some $SU(3)$ CGC not
explicitly listed there can be obtained by the symmetry properties from
the listed ones.  Table 2 gives the additional needed phase factors
$\epsilon_2$ which are missing in table 5 of Sec.\ II of ref.[8b].)

It is time consuming and requires a good grasp of group theory to
combine the individual isoscalar factors into the transformation
coefficients between physical bases and symmetry bases and the $6
\rightarrow 4 + 2$ cfp for the matrix element calculations.

In order to make the calculation automatic and to facilitate others
using this $fp$ expansion technique, a computer program has been
written.  All the needed isoscalar factors are stored in the program.
After inputting the quantum numbers $\alpha = (YIJ)$, the program will
automatically yield the physical bases, symmetry bases, the
transformation coefficients between these two bases and the $6
\rightarrow 4 + 2$ cfp for the symmetry bases.  This part may be useful
for other dibaryon model practitioners if they want to use the $fp$
expansion methods.  For our own problem, the program continues on to
calculate the one body, two body matrix elements, the four body
overlap, combine them together into the six quark Hamiltonian matrix
elements in the physical bases, diagonalize the Hamiltonian in the
non-orthogonal physical basis space, minimize the eigen-energy and fix
the eigen-WF with respect to the delocalization parameter
$\epsilon(s)$, and repeat this calculation for different separations
$s$ between two $q^3$ clusters from $s =$ 0.1 to 3 fm, and finally
outputs the adiabatic potential $V_{\alpha}(s)$.  This program greatly
reduced the labor involved in the systematic search of dibaryon
candidates in the $u,d$ and $s$ three flavor world.  Only minor
modification of the subroutine for the one and two body matrix elements
calculation, suffices to adapt the program to a relativistic quark
model dibaryon search.  We expect it is also easy to apply this program
to other nonrelativistic and relativistic dibaryon calculations with
minor modifications, particularly as the fp expansion part is universal
for this kind of dibaryon model calculations.

\begin{center}
\large{\bf ACKNOWLEDGEMENTS}
\end{center}

Enlightening discussions with Prof.\ J.Q.\ Chen are gratefully
acknowledged.

This work is supported by the NSFC, the fundamental research fund of
State Science and Technology Committee of China, the graduate study
fund of State Educational Committee of China and the Department of
Energy of the United States.

\begin{center}
\large{\bf REFERENCES}
\end{center}

\begin{enumerate}

\item [[1]]  T.S.\ Walhout and J.\ Wambach, Phys.\ Rev.\ Lett.\ {\bf
67}, 314 (1991); N.R.\ Walet, R.D.\ Amado, and A.\ Hosaka,
Phys.\ Rev.\ Lett.\ {\bf 68}, 3849 (1992).

\item [[2]]  F.\ Wang, G.H.\ Wu, L.J.\ Teng, and T.\ Goldman,
Phys.\ Rev.\ Lett.\ {\bf 69}, 2901 (1992).  G.H.\ Wu, L.J.\ Teng,
J.L.\ Ping, F.\ Wang, and T.\ Goldman, Los Alamos preprint
LA-UR-94-2326.

\item [[3]]  R.L.\ Jaffe, Phys.\ Rev.\ Lett.\ {\bf 38}, 195 (1977).

\item [[4]]  B.\ Silvestre-Brac and J.\ Leandri, Phys.\ Rev.\ {\bf D
45}, 4221 (1992); {\bf D 47}, 5083 (1993), and references therein.

\item [[5]]  F.\ Wang and Y.\ He, Chin.\ J.\ Nucl.\ Phys.\ {\bf 4}, 176
(1982); {\bf 5}, 263 (1983); Prog.\ Phys., {\bf 4}, 34 (1984), (in
Chinese); N.\ Isgur and J.\ Paton, Phys.\ Rev.\ {\bf D 31}, 283 (1985);
H.\ Markum {\it et al.}, Phys.\ Rev.\ {\bf D 31}, 2029 (1985);
S.\ Ohta, M.\ Fukugita and A.\ Ukawa, Phys.\ Lett.\ {\bf B 173}, 15
(1986); Yu.A.\ Simonov, Phys.\ Lett.\ {\bf B 228}, 413 (1989).

\item [[6]]  T.\ Goldman, K.\ Maltman, G.J.\ Stephenson Jr., K.E.
Schmidt, and F.\ Wang, Phys.\ Rev.\ Lett.\ {\bf 59}, 627 (1987);
Phys.\ Rev.\ {\bf C 39}, 1889 (1989).

\item [[7]]  F.\ Wang, J.L.\ Ping, G.H.\ Wu, L.J.\ Teng, and
T.\ Goldman, Los Alamos preprint LA-UR-94-2363 (submitted together).

\item [[8a]]  J.Q.\ Chen, J.\ Math.\ Phys.\ {\bf 22}, 1 (1981);
J.Q.\ Chen, {\it Group Representation Theory for Physicists}, 1989
(World Scientific, Singapore).

\item [[8b]]  J.Q.\ Chen {\it et al.}, {\it Tables of the
Clebsh-Gordan, Racah and Subduction Coefficients of} $SU(n)$ {\it
Groups,} 1987 (World Scientific, Singapore); {\it Tables of the}
$SU(mn) \supset SU(m) \times SU(n)$ {\it Coefficients of Fractional
Parentage,} 1991 (World Scientific, Singapore).

\item [[9]]  M.\ Harvey, Nucl.\ Phys.\ {\bf A 352}, 301 (1981); {\bf A
481}, 834 (1988).

\item [[10]]  J.Q. Chen {\it et al.}, Nucl.\ Phys. {\bf A 393}, 122
(1983).

\item [[11]]  Y.C.\ Tang, M.\ LaMere and D.R.\ Thompson,
Phys.\ Rep.\ {\bf 47C}, 169 (1978).

\item [[12]]  F.\ Wang and C.W.\ Wong, Nuovo Cimento, {\bf A 86}, 283
(1985); Prog.\ Phys.\ {\bf 9}, 297 (1989); in {\it Quark-gluon
Structure of Hadron and Nuclei,} ed.\ L.S.\ Kisslinger and S.J.\ Qiu
(International Academic Press, 1991) p.\ 100; H.\ Hofestadt, S.\ Merk,
and H.R.\ Petry, Z.\ Phys.\ {\bf A 326}, 391 (1987); R.L.\ Jaffe,
Nucl.\ Phys.\ {\bf A 522}, 365c (1991).

\item [[13]]  C.R.\ Dover, Nucl.\ Phys.\ {\bf A 450}, 95c (1986);
M.\ Oka, Phys.\ Rev.\ {\bf D 38}, 298 (1988); K.\ Shimizu,
Rep.\ Prog.\ Phys.\ {\bf 52}, 1 (1989).

\item [[14]]  C.L.\ Critchfield, Phys.\ Rev.\ {\bf D 12}, 923 (1975);
J.\ Math.\ Phys.\ {\bf 17}, 261 (1976).

\item [[15]]  Fl.\ Stancu and L.\ Wilets, Phys.\ Rev.\ {\bf C 40}, 1901
(1989).

\item [[16]] T.\ Goldman, K.\ Maltman, G.J.\ Stephenson Jr., and K.E.
Schmidt, Nucl.\ Phys.\ {\bf A481}, 621 (1988).

\end{enumerate}

\pagebreak

\begin{center}
\large{\bf APPENDIX}
\end{center}

Our Tables for the symmetry decompositions might appear to contradict
Harvey's results [9].  This is due to a difference in terminology:
Harvey used `symmetric' and `antisymmetric' to refer only to the
\underline{orbital} components when discussing non-identical
particles.  We prefer to use the more inclusive definition below, since
we believe it allows for a more natural relation to the identical
particle case.

The symmetric (antisymmetric) combination $\bar{xy}
(\stackrel{\sim}{xy})$ of two baryon state is defined as
$$
\bar{xy} = \frac{xy + yx}{\sqrt{2}}, \hspace*{0.250in}
\stackrel{\sim}{xy} = \frac{xy - yx}{\sqrt{2}} \eqno (A1)
$$
Let's use the $N \Delta$ two baryon state as an example to show the
symmetry property.  Below $\chi_c$ is the color singlet three quark
state, $N_{m_{N} \tau_{N}} (\Delta_{m_{\Delta} \tau_{\Delta}})$ is a
three quark $N (\Delta)$ spin-isospin symmetric state with spin-isospin
projection quantum numbers $m_{N} \tau_{N} (m_{\Delta} \tau_{\Delta})$,
$l (123)$ is a product orbital state $l(1) l(2) l(3)$ and $l$ is
defined in eq.\ (4), $r (456)$ has the parallel meaning, $C^{Kk}_{K_{a}
k_{a}, K_{b} k_{b}}$ is the spin (isospin) CGC.  Then
$$
(N \Delta)_{IJ} = A \Sigma C^{J m}_{J_{N} m_{N}, J_{\Delta} m_{\Delta}}
C^{I \tau}_{I_{N} \tau_{N}, I_{\Delta} \tau_{\Delta}}
$$
$$
\chi_c (123) N_{m_{N} \tau_{N}}(123) l (123) \chi_c (456)
\Delta_{m_{\Delta} \tau_{\Delta}} (456) r (456)
$$
$$
= \frac{1}{\sqrt{20}} \Sigma C^{J m}_{J_{N} m_{N}, J_{\Delta}
m_{\Delta}} C^{I \tau}_{I_{N} \tau_{N}, I_{\Delta} \tau_{\Delta}}
\lbrace \chi_c (123) \chi_c (456)
$$
$$
\left[ N_{m_{N} \tau_{N}} (123) \Delta_{m_{\Delta}\tau_{\Delta}} (456)
l(123) r(456) \right.
$$
$$
\left. - N_{m_{N} \tau_{N}} (456) \Delta_{m_{\Delta} \tau_{\Delta}}
(123) l(456) r(123) \right] + .... \rbrace \eqno (A2)
$$
$$
(\Delta N)_{IJ} = A \Sigma C^{J m}_{J_{\Delta} m_{\Delta}, J_{N} m_{N}}
C^{I \tau}_{I_{\Delta} \tau_{\Delta}, I_{N} \tau_{N}}
$$
$$
\chi_c (123) \Delta_{m_{\Delta} \tau_{\Delta}} (123) l(123) \chi_c
(456) N_{m_{N} \tau_{N}} (456) r(456)
$$
$$
= \frac{1}{\sqrt{20}} \Sigma C^{Jm}_{J_{\Delta} m_{\Delta}, J_{N}
m_{N}} C^{\; I_{\tau} \;}_{I_{\Delta} \tau_{\Delta}, I_{N} \tau_{N}}
\lbrace \chi_c (123) \chi_c(456)
$$
$$
\left. \left[ \Delta_{m_{\Delta} \tau_{\Delta}} (123) N_{m_{N}
\tau_{N}} (456) l(123) r(456) \right. \right.
$$
$$
\left. - \Delta_{m_{\Delta} \tau_{\Delta}}(456) N_{m_{N} \tau_{N}}(123)
l(456) r(123) \right] + .... \rbrace \eqno (A.3)
$$
where $+ ...$ represents all of the other permutations.

$$
(\overline{N \Delta})_{IJ} = \frac{(N \Delta)_{IJ} + (\Delta
N)_{IJ}}{\sqrt{2}}
$$
$$
= \frac{1}{\sqrt{40}} \Sigma C^{Jm}_{J_{N} m_{N}, J_{\Delta}
m_{\Delta}} C^{I \tau}_{I_{N} \tau_{N}, I_{\Delta}
\tau_{\Delta}}\lbrace \chi_c (123) \chi_c (456)
$$
$$
\left[ N_{m_{N} \tau_{N}} (123) \Delta_{m_{\Delta} \tau_{\Delta}} (456)
\left( l(123) r(456) - (-)^{J_N + J_{\Delta} - J + I_N + I_{\Delta} -I}
l(456) r(123) \right) \right.
$$
$$
\left. - N_{m_{N} \tau_{N}} (456) \Delta_{m_{\Delta} \tau_{\Delta}}
(123) \left( l(456) r(123) - (-)^{J_N + J_{\Delta} - J + I_N +
I_{\Delta} -I} l(123) r(456) \right) \right] + .... \rbrace \eqno (A.4)
$$
The orbital symmetry property (the parity) of a two baryon state under
the permutation $\left(^{123, 456}_{456, 123} \right)$ is dependent on
the spin-isospin quantum numbers, instead of directly related to the
symmetry (antisymmetry) $\bar{xy} (\stackrel{\sim}{xy})$ combination as
explained in [9].

In deriving (A.4), we have used the well known $SU(2)$ relation $$
C^{K_k}_{K_a k_a K_b k_b} = (-)^{K_a + K_b -K} C^{Kk}_{K_b k_b K_a k_a}
\eqno (A.5)
$$
Note, for example, that if the $\Delta$ were replaced by a second $N$,
and $I + J$ is even, the first and fourth terms become identical as do
the second and third terms, etc.  The result has \underline{only}
antisymmetric orbital parts.  The $(\overline{NN})_{IJ} \equiv 0$ for
odd $I + J$.  Conversely, for $(\overline{NN})_{IJ}$, only the odd $I +
J$ symmetric orbital parts exist, (as for example, in the deuteron).

\pagebreak

\begin{center}
\large{\bf TABLE CAPTIONS}
\end{center}

\begin{enumerate}

\item TABLE 1a.  The transformation coefficients between physical bases
and symmetry bases for Y = 2.  The head of columns is $[\nu]$, $[\mu]$,
$[f]$: 1 - [6]; 2 - [51]; 3 - [42]; 4 - [33]; 5 - [411]; 6 - [321]; 7 -
[222].  The head of rows is $B_1 B_2$: 1 - N; 2 -$\Sigma$; 3 - $\Xi$; 4
- $\Lambda$; 5 - $\Delta$; 6 - $\Sigma^{*}$; 7 - $\Xi^{*}$; 8 -
$\Omega$.  $\bar{xy} (\stackrel{\sim}{xy})$ means symmetric
(antisymmetric) channel of baryons $x$ and $y$.  The transformation
coefficients should be the square root of the entries, and a negative
sign means to take the negative square root.

\item TABLE 1b.  Same as table 1a.\ for Y = 1.

\item TABLE 1c.  Same as table 1a.\ for Y = 0.

\item TABLE 1d.  Same as table 1a.\ for Y = -1.

\item TABLE 1e.  Same as table 1a.\ for Y = -2.

\item TABLE 1f.  Same as table 1a.\ for Y = -3.

\item TABLE 1g.  Same as table 1a.\ for Y = -4.

\item TABLE 2.  The additional phase factor $\epsilon_2 (\nu_1 \nu_2
\nu)$.

\end{enumerate}

\pagebreak

\vspace*{-1.75in}

\begin{center}
{\bf TABLE 1a}\\
\vspace*{0.05in}
\small{
\begin{tabular}{p{45pt}|p{45pt}p{45pt}p{45pt}p{45pt}p{45pt}p{45pt}}
\hline
{IJ = 33} & {411} \\ \hline
55 & 1 \\ \hline
\hline
IJ = 32 & 321 \\ \hline
55 & 1 \\ \hline
\hline
IJ = 31 & 231 & 431 \\ \hline
55 & -5/9 & -4/9 \\ \hline
\hline
IJ = 30 & 141 & 341 \\ \hline
55 & -1/5 & -4/5 \\ \hline
\hline
IJ = 23 & 322 \\ \hline
55 & 1 \\ \hline
\hline
IJ = 22 & 232 & 322 & 412 & 432 \\ \hline
$\overline{15}$ & 1/9 & 0 & 4/5 & 4/45 \\
$\widetilde{51}$ & 0 & -1 & 0 & 0 \\
55 & -4/9 & 0 & 1/5 & -16/45 \\ \hline
\hline
IJ = 21 & 142 & 232 & 322 & 342 & 432 \\ \hline
$\overline{15}$ & 4/45 & 0 & 5/9 & 16/45 & 0 \\
$\widetilde{15}$ & 0 & 5/9 & 0 & 0 & 4/9 \\
55 & -1/9 & 0 & 4/9 & -4/9 & 0 \\ \hline
\hline
IJ = 20 & 232 & 432 \\ \hline
55 & -5/9 & -4/9 \\ \hline
\hline
IJ = 13 & 233 & 433 \\ \hline
55 & -5/9 & -4/9 \\ \hline
\hline
IJ = 12 & 143 & 233 & 323 & 343 & 4133 \\ \hline
$\overline{15}$ & 4/45 & 0 & 5/9 & 16/45 & 0 \\
$\widetilde{15}$ & 0 & 5/9 & 0 & 0 & 4/9 \\
55 & -1/9 & 0 & 4/9 & -4/9 & 0 \\ \hline
\hline
IJ = 11 & 2$_1$33 & 2$_2$33 & 323 & 413 & 4$_1$33 & 4$_2$33 \\ \hline
11 & 5/81 & 20/81 & 0 & 4/9 & 4/81 & 16/81 \\
$\overline{15}$ & -20/81 & -5/81 & 0 & 4/9 & -16/81 & -4/81 \\
$\widetilde{15}$ & 0 & 0 & -1 & 0 & 0 & 0 \\
55 & 20/81 & -20/81 & 0 & 1/9 & 16/81 & -16/81 \\ \hline
\hline
IJ = 10 & 143 & 323 & 343 \\ \hline
11 & 1/9 & 4/9 & 4/9 \\
55 & -4/45 & 5/9 & -16/45 \\ \hline
\hline
IJ = 03 & 144 & 344 \\ \hline
55 & -1/5 & -4/5 \\ \hline
\hline
IJ = 02 & 234 & 434 \\ \hline
55 & -5/9 & -4/9 \\ \hline
\hline
IJ = 01 & 144 & 324 & 344 \\ \hline
11 & 1/9 & 4/9 & 4/9 \\
55 & -4/45 & 5/9 & -16/45 \\ \hline
\hline
IJ = 00 & 234 & 414 & 434 \\ \hline
11 & 1/9 & 4/5 & 4/45 \\
55 & -4/9 & 1/5 & -16/45 \\ \hline
\end{tabular}}
\end{center}

\pagebreak
\vspace*{-1.75in}
\begin{center}
{\bf TABLE 1b}\\
\vspace*{0.05in}
\hspace*{-0.5in}
\small{
\begin{tabular}{p{30pt}|p{30pt}p{35pt}p{35pt}p{40pt}p{46pt}p{35pt}p{35pt}
p{35pt}p{35pt}p{35pt}}
\hline
{IJ=$\frac{5}{2}$3} & {322} & {411}\\ \hline
$\overline{56}$ & 0 & 1 \\ \hline
$\widetilde{56}$ & 1 & 0 \\ \hline
\hline
IJ=$\frac{5}{2}$2 & 232 & 321 & 322 & 412 & 432 \\ \hline
$\overline{25}$ & 1/9 & 0 & 0 & 4/5 & 4/45 \\
$\widetilde{25}$ & 0 & 0 & -1 & 0 & 0 \\
$\overline{56}$ & 0 & 1 & 0 & 0 & 0 \\
$\widetilde{56}$ & -4/9 & 0 & 0 & 1/5 & -16/45 \\ \hline
\hline
IJ=$\frac{5}{2}$1 & 142 & 231 & 232 & 322 & 342 & 431 & 432 \\ \hline
$\overline{25}$ & 4/45 & 0 & 0 & 5/9 & 16/45 & 0 & 0 \\
$\widetilde{25}$ & 0 & 0 & 5/9 & 0 & 0 & 0 & 4/9 \\
$\overline{56}$ & 0 & -5/9 & 0 & 0 & 0 & -4/9 & 0 \\
$\widetilde{56}$ & -1/9 & 0 & 0 & 4/9 & -4/9 & 0 & 0 \\ \hline
\hline
IJ=$\frac{3}{2}0$ & 141 & 232 & 341 & 432 \\ \hline
$\overline{56}$ & -1/5 & 0 & -4/5 & 0 \\
$\widetilde{56}$ & 0 & -5/9 & 0 & -4/9 \\ \hline
IJ=$\frac{3}{2}$3 & 233 & 322 & 433 \\ \hline
$\overline{56}$ & 0 & 1 & 0 \\
$\widetilde{56}$ & -5/9 & 0 & -4/9 \\ \hline
\hline
IJ=$\frac{3}{2}2$ & 143 & 232 & 233 & 235 & 322 & 323 & 325 & 343 &
412 & 432 \\ \hline
$\overline{16}$ & 0 & 5/72 & 5/72 & -5/36 & 0 & 0 & 0 & 0 & 1/2 & 1/18 \\
$\overline{25}$ & 1/36 & 0 & 0 & 0 & -1/16 & 25/144 & -5/8 & 1/9 & 0 & 0 \\
$\overline{45}$ & 0 & 5/144 & -5/16 & 5/72 & 0 & 0 & 0
& 0 & 1/4 & 1/36 \\
$\widetilde{25}$ & 0 & 1/144 & 25/144 & 25/72 & 0 & 0 & 0 & 0 & 1/20 &
1/180 \\
$\widetilde{15}$ & -1/20 & 0 & 0 & 0 & -5/16 & -5/16 & -1/8 & -1/5 & 0 & 0
\\
$\overline{56}$ & 0 & -4/9 & 0 & 0 & 0 & 0 & 0 & 0 & 1/5 & -16/45 \\
$\widetilde{16}$ & 1/90 & 0 & 0 & 0 & -5/8 & 5/72 & 1/4 & 2/45 & 0 & 0 \\
$\widetilde{56}$ & -1/9 & 0 & 0 & 0 & 0 & 4/9 & 0 & -4/9 & 0 & 0 \\
\hline
& 433 & 435 \\ \hline
$\overline{16}$ & 1/18 & -1/9 \\
$\overline{25}$ & 0 & 0 \\
$\overline{45}$ & -1/4 & 1/18 \\
$\widetilde{25}$ & 5/36 & 5/18 \\
$\widetilde{45}$ & 0 & 0 \\
$\overline{56}$ & 0 & 0 \\
$\widetilde{16}$ & 0 & 0 \\
$\widetilde{56}$ & 0 & 0 \\ \hline
\end{tabular}}
\end{center}

\pagebreak
\vspace*{-1.75in}
\begin{center}
{\bf TABLE 1b Cont-}\\
\vspace*{0.05in}
\hspace*{-1.0in}
\small{
\begin{tabular}{p{30pt}|p{33pt}p{35pt}p{35pt}p{40pt}p{46pt}p{35pt}p{38pt}
p{35pt}p{35pt}p{35pt}p{35pt}}
\hline
IJ=$\frac{3}{2}1$ & 142 & 145 & 232 & 23$_1$3 & 23$_2$3 & 235 & 322
& 323 & 325 & 342 \\ \hline
$\overline{16}$ & 1/18 & -2/45 & 0 & 0 & 0 & 0 & 25/72 & -1/8 & 1/36 & 2/9
\\
$\overline{25}$ & 0 & 0 & 5/144 & -25/324 & -25/1296 & 25/72 & 0 & 0 & 0
& 0 \\
$\overline{45}$ & 1/36 & 1/45 & 0 & 0 & 0 & 0 & 25/144 & 9/16 & -1/72 &
1/9 \\
$\widetilde{25}$ & 1/180 & 1/9 & 0 & 0 & 0 & 0 & 5/144 & -5/16 & -5/72 &
1/45 \\
$\widetilde{45}$ & 0 & 0 & 25/144 & 5/36 & 5/144 & 5/72 & 0 & 0 & 0 & 0 \\
$\overline{56}$ & -1/9 & 0 & 0 & 0 & 0 & 0 & 4/9 & 0 & 0 & -4/9 \\
$\widetilde{16}$ & 0 & 0 & 25/72 & -5/162 & -5/648 & -5/36 & 0 & 0 & 0 &
0 \\
$\widetilde{56}$ & 0 & 0 & 0 & 20/81 & -20/81 & 0 & 0 & 0 & 0 & 0 \\
$\overline{12}$ & 0 & 0 & 0 & 5/81 & 20/81 & 0 & 0 & 0 & 0 & 0 \\
$\widetilde{12}$ & 0 & 1/45 & 0 & 0 & 0 & 0 & 0 & 0 & 8/9 & 0 \\ \hline
& 345 & 413 & 432 & 43$_1$3 & 43$_2$3 & 435 \\ \hline
$\overline{16}$ & -3/45 & 0 & 0 & 0 & 0 & 0 \\
$\overline{25}$ & 0 & 5/36 & 1/36 & -5/81 & -5/324 & 5/18 \\
$\overline{45}$ & 4/45 & 0 & 0 & 0 & 0 & 0 \\
$\widetilde{25}$ & 4/9 & 0 & 0 & 0 & 0 & 0 \\
$\widetilde{45}$ & 0 & -1/4 & 5/36 & 1/9 & 1/36 & 1/18 \\
$\overline{56}$ & 0 & 0 & 0 & 0 & 0 & 0 \\
$\widetilde{16}$ & 0 & 1/18 & 5/18 & -2/81 & -1/162 & -1/9 \\
$\widetilde{56}$ & 0 & 1/9 & 0 & 16/81 & -16/81 & 0 \\
$\overline{12}$ & 0 & 4/9 & 0 & 4/81 & 16/81 & 0 \\
$\widetilde{12}$ & 4/45 & 0 & 0 & 0 & 0 & 0 \\ \hline
\hline
IJ=$\frac{3}{2}$0 & 143 & 232 & 235 & 323 & 343 & 432 & 435 \\ \hline
$\overline{12}$ & 1/9 & 0 & 0 & 4/9 & 4/9 & 0 & 0 \\
$\widetilde{12}$ & 0 & 0 & -5/9 & 0 & 0 & 0 & -4/9 \\
$\overline{56}$ & 0 & -5/9 & 0 & 0 & 0 & -4/9 & 0 \\
$\widetilde{56}$ & -4/45 & 0 & 0 & 5/9 & -16/45 & 0 & 0 \\ \hline
\hline
IJ=$\frac{1}{2}$3 & 144 & 233 & 344 & 433 \\ \hline
$\overline{56}$ & 0 & -5/9 & 0 & -4/9 \\
$\widetilde{56}$ & -1/5 & 0 & -4/5 & 0 \\ \hline
\hline
IJ=$\frac{1}{2}$2 & 143 & 146 & 233 & 234 & 236 & 323 & 343 & 346 &
433 & 434 & 436 \\ \hline
$\overline{16}$ & 16/255 & -1/25 & 0 & 0 & 0 & 4/9 & 64/225 & -4/25 & 0 &
0 & 0 \\
$\overline{25}$ & 0 & 0 & 1/9 & 0 & 4/9 & 0 & 0 & 0 & 4/45 & 0 & 16/45 \\
$\widetilde{25}$ & 4/225 & 4/25 & 0 & 0 & 0 & 1/9 & 16/225 & 16/25 & 0 &
0 & 0 \\
$\overline{56}$ & -1/9 & 0 & 0 & 0 & 0 & 4/9 & -4/9 & 0 & 0 & 0 & 0 \\
$\widetilde{16}$ & 0 & 0 & 4/9 & 0 & -1/9 & 0 & 0 & 0 & 16/45 & 0 & -4/45
\\
$\widetilde{56}$ & 0 & 0 & 0 & -5/9 & 0 & 0 & 0 & 0 & 0 & -4/9 & 0 \\ \hline
\hline
IJ=$\frac{1}{2}$1 & 144 & 146 & 23$_1$3 & 23$_2$3 & 23$_1$6 &
23$_2$6 & 323 & 324 & 326 & 344 \\ \hline
$\overline{12}$ & 1/18 & 2/45 & 0 & 0 & 0 & 0 & 0 & 2/9 & -5/18 & 2/9 \\
$\overline{14}$ & 0 & 0 & 1/18 & 2/9 & -1/36 & -1/36 & 0 & 0 & 0 & 0 \\
$\widetilde{12}$ & 0 & 0 & -1/162 & -2/81 & -1/4 & -1/4 & 0 & 0 & 0 & 0 \\
$\widetilde{14}$ & 1/18 & -2/25 & 0 & 0 & 0 & 0 & 0 & 2/9 & 5/18 & 2/9 \\
$\overline{56}$ & 0 & 0 & 20/81 & -20/81 & 0 & 0 & 0 & 0 & 0 & 0 \\
$\widetilde{56}$ & -4/45 & 0 & 0 & 0 & 0 & 0 & 0 & 5/9 & 0 & -16/45 \\
$\overline{16}$ & 0 & 0 & -16/81 & -4/81 & -1/18 & 1/18 & 0 & 0 & 0 & 0 \\
$\overline{25}$ & 0 & -4/45 & 0 & 0 & 0 & 0 & -1/5 & 0 & -16/45 & 0 \\
$\widetilde{25}$ & 0 & 0 & -4/81 & -1/81 & 2/9 & -2/9 & 0 & 0 & 0 & 0 \\
$\widetilde{16}$ & 0 & 1/45 & 0 & 0 & 0 & 0 & -4/5 & 0 & 4/45 & 0 \\ \hline
\end{tabular}}
\end{center}

\pagebreak
\vspace*{-1.1in}
\begin{center}
{\bf TABLE 1b Cont-}\\
\vspace*{0.05in}
\hspace*{-0.25in}
\small{
\begin{tabular}{p{30pt}|p{40pt}p{35pt}p{40pt}p{40pt}p{46pt}p{35pt}p{35pt}
p{35pt}p{35pt}p{35pt}p{35pt}}
\hline
& 346 & 413 & 43$_1$3 & 43$_2$3 & 43$_1$6 & 43$_2$6 \\ \hline
$\overline{12}$ & 8/45 & 0 & 0 & 0 & 0 & 0 \\
$\overline{14}$ & 0 & 2/5 & 2/45 & 8/45 & -1/45 & -1/45 \\
$\widetilde{12}$ & 0 & -2/45 & -2/405 & -8/405 & -1/5 & -1/5 \\
$\widetilde{14}$ & -8/45 & 0 & 0 & 0 & 0 & 0 \\
$\overline{56}$ & 0 & 1/9 & 16/81 & -16/81 & 0 & 0 \\
$\widetilde{56}$ & 0 & 0 & 0 & 0 & 0 & 0 \\
$\overline{16}$ & 0 & 16/45 & -64/405 & -16/405 & -2/45 & 2/45 \\
$\overline{25}$ & -16/45 & 0 & 0 & 0 & 0 & 0 \\
$\widetilde{25}$ & 0 & 4/45 & -16/405 & -4/405 & 8/45 & -8/45 \\
$\widetilde{16}$ & 4/45 & 0 & 0 & 0 & 0 & 0 \\ \hline
\hline
IJ=$\frac{1}{2}$0 & 143 & 234 & 236 & 323 & 326 & 343 & 414 & 434 &
436 \\ \hline
$\overline{12}$ & 0 & 1/18 & 5/18 & 0 & 0 & 0 & 2/5 & 2/45 & 2/9 \\
$\overline{14}$ & 1/10 & 0 & 0 & 2/5 & 1/10 & 2/5 & 0 & 0 & 0 \\
$\widetilde{12}$ & -1/90 & 0 & 0 & -2/45 & 9/10 & -2/45 & 0 & 0 & 0 \\
$\widetilde{14}$ & 0 & 1/18 & -5/18 & 0 & 0 & 0 & 2/5 & 2/45 & -2/9 \\
$\overline{56}$ & -4/45 & 0 & 0 & 5/9 & 0 & -16/45 & 0 & 0 & 0 \\
$\widetilde{56}$ & 0 & -4/9 & 0 & 0 & 0 & 0 & 1/5 & -16/45 & 0 \\
\hline
\end{tabular}}
\end{center}

\pagebreak
\vspace*{-1.1in}
\begin{center}
{\bf TABLE 1c}\\
\end{center}
\vspace*{0.05in}
\hspace*{-0.50in}
\small{\begin{tabular}{p{35pt}|p{35pt}p{40pt}p{40pt}p{35pt}p{35pt}p{35pt}p
{35pt}p{35pt}p{35pt}p{35pt}}
\hline
IJ=23 & 233 & 322 & 411 & 433 \\ \hline
$\overline{57}$ & -1/3 & 0 & 2/5 & -4/15 \\
66 & 2/9 & 0 & 3/5 & 8/45 \\
$\widetilde{57}$ & 0 & 1 & 0 & 0 \\ \hline
\hline
IJ=22 & 143 & 232 & 233 & 321 & 322 & 323 & 343 & 412 & 432 & 433 \\
\hline
$\overline{26}$ & 0 & 1/12 & 5/36 & 0 & 0 & 0 & 0 & 3/5 & 1/15 & 1/9 \\
$\overline{35}$ & 0 & 1/36 & -5/12 & 0 & 0 & 0 & 0 & 1/5 & 1/45 & -1/3 \\
$\widetilde{35}$ & -1/15 & 0 & 0 & 0 & -1/4 & -5/12 & -4/15 & 0 & 0 & 0 \\
$\overline{57}$ & -1/15 & 0 & 0 & 2/5 & 0 & 4/15 & -4/15 & 0 & 0 & 0 \\
$\widetilde{26}$ & 1/45 & 0 & 0 & 0 & -3/4 & 5/36 & 4/45 & 0 & 0 & 0 \\
66 & 2/45 & 0 & 0 & 3/5 & 0 & -8/45 & 8/45 & 0 & 0 & 0 \\
$\widetilde{57}$ & 0 & -4/9 & 0 & 0 & 0 & 0 & 0 & 1/5 & -16/45 & 0 \\ \hline
\hline
IJ=21 & 142 & 231 & 232 & 23$_1$3 & 23$_2$3 & 322 & 323 & 342 &
413 & 431 \\ \hline
$\overline{26}$ & 1/15 & 0 & 0 & 0 & 0 & 5/12 & -1/4 & 4/15 & 0 & 0 \\
$\overline{35}$ & 1/45 & 0 & 0 & 0 & 0 & 5/36 & 3/4 & 4/45 & 0 & 0 \\
$\widetilde{35}$ & 0 & 0 & 5/36 & 5/27 & 5/108 & 0 & 0 & 0 & -1/3 & 0 \\
$\overline{57}$ & 0 & -2/9 & 0 & 4/27 & -4/27 & 0 & 0 & 0 & 1/15 & -8/45 \\
$\widetilde{26}$ & 0 & 0 & 5/12 & -5/81 & -5/324 & 0 & 0 & 0 & 1/9 & 0 \\
$\widetilde{57}$ & -1/9 & 0 & 0 & 0 & 0 & 4/9 & 0 & -4/9 & 0 & 0 \\
66 & 0 & -1/3 & 0 & -8/81 & 8/81 & 0 & 0 & 0 & -2/45 & -4/15 \\
22 & 0 & 0 & 0 & 5/81 & 20/81 & 0 & 0 & 0 & 4/9 & 0 \\ \hline
& 432 & 43$_1$3 & 43$_2$3 \\ \hline
$\overline{26}$ & 0 & 0 & 0 \\
$\overline{35}$ & 0 & 0 & 0 \\
$\widetilde{35}$ & 1/9 & 4/27 & 1/27 \\
$\overline{57}$ & 0 & 16/135 & -16/135 \\
$\widetilde{26}$ & 1/3 & -4/81 & -1/81 \\
$\widetilde{57}$ & 0 & 0 & 0 \\
66 & 0 & -32/405 & 32/405 \\
22 & 0 & 4/31 & 16/31 \\ \hline
\hline
IJ=20 & 141 & 143 & 232 & 323 & 341 & 343 & 432 \\ \hline
$\overline{57}$ & -2/25 & -4/75 & 0 & 1/3 & -8/25 & -16/75 & 0 \\
66 & -3/25 & 8/225 & 0 & -2/9 & -12/25 & 32/225 & 0 \\
$\widetilde{57}$ & 0 & 0 & -5/9 & 0 & 0 & 0 & -4/9 \\
22 & 0 & 1/9 & 0 & 4/9 & 0 & 4/9 & 0 \\ \hline
\end{tabular}}

\pagebreak
\vspace*{-1.1in}
\begin{center}
{\bf TABLE 1c Cont-}\\
\end{center}
\vspace*{0.05in}
\hspace*{-0.50in}
\small{\begin{tabular}{p{35pt}|p{35pt}p{35pt}p{35pt}p{40pt}p{35pt}p{35pt}p
{35pt}p{40pt}p{35pt}p{35pt}}
\hline
IJ=13 & 144 & 233 & 322 & 344 & 433 \\ \hline
$\overline{57}$ & -2/15 & 0 & 1/3 & -3/15 & 0 \\
66 & 1/15 & 0 & 2/3  & 4/15 & 0\\
$\widetilde{57}$ & 0 & 5/9 & 0 & 0 & -4/9\\ \hline
\hline
IJ=12 & 143 & 146 & 232 & 233 & 234 & 235 & 236 & 322 & 323 & 325 \\
\hline
$\overline{17}$ & 0 & 0 & 1/27 & 1/9 & 0 & -5/27 & -2/27 & 0 & 0 & 0 \\
$\overline{26}$ & 1/25 & 2/75 & 0 & 0 & 0 & 0 & 0 & -1/12 & 1/4 & -1/3 \\
$\overline{35}$ & -1/225 & -8/75 & 0 & 0 & 0 & 0 & 0 & -1/12 & -1/36 &
-1/3 \\
$\overline{46}$ & 0 & 0 & 1/18 & -1/6 & 0 & 0 & 1/9 & 0 & 0 & 0 \\
$\widetilde{35}$ & 0 & 0 & 1/108 & -1/36 & 0 & 5/27 & -8/27 & 0 & 0 & 0 \\
$\overline{57}$ & 0 & 0 & -4/27 & 0 & -10/27 & 0 & 0 & 0 & 0 & 0 \\
$\widetilde{26}$ & 0 & 0 & 1/108 & 1/4 & 0 & 5/27 & 2/27 & 0 & 0 & 0 \\
$\widetilde{46}$ & -2/75 & 1/25 & 0 & 0 & 0 & 0 & 0 & -1/2 & -1/6 & 0 \\
$\widetilde{17}$ & 4/225 & -2/75 & 0 & 0 & 0 & 0 & 0 & -1/3 & 1/9 & 1/3 \\
$\widetilde{57}$ & -1/9 & 0 & 0 & 0 & 0 & 0 & 0 & 0 & 4/9 & 0 \\
66 & 0 & 0 & -8/27 & 0 & 5/27 & 0 & 0 & 0 & 0 & 0 \\ \hline
& 343 & 346 & 412 & 432 & 433 & 434 & 435 & 436 \\ \hline
$\overline{17}$ & 0 & 0 & 4/15 & 4/135 & 4/45 & 0 & -4/27 & -8/135 \\
$\overline{26}$ & 4/25 & 8/75 & 0 & 0 & 0 & 0 & 0 & 0 \\
$\overline{35}$ & -4/225 & -32/75 & 0 & 0 & 0 & 0 & 0 & 0 & \\
$\overline{46}$ & 0 & 0 & 2/5 & 2/45 & -2/15 & 0 & 0 & 4/45 \\
$\widetilde{35}$ & 0 & 0 & 1/15 & 1/135 & -1/45 & 0 & 4/27 & -32/135 \\
$\overline{57}$ & 0 & 0 & 1/15 & -16/135 & 0 & -8/27 & 0 & 0 \\
$\widetilde{26}$ & 0 & 0 & 1/15 & 1/135 & 1/5 & 0 & 4/27 & 8/135 \\
$\widetilde{46}$ & -8/75 & 4/25 & 0 & 0 & 0 & 0 & 0 & 0 \\
$\widetilde{17}$ & 16/225 & -8/75 & 0 & 0 & 0 & 0 & 0 & 0 \\
$\widetilde{57}$ & -4/9 & 0 & 0 & 0 & 0 & 0 & 0 & 0 \\
66 & 0 & 0 & 2/15 & -32/135 & 0 & 4/27 & 0 & 0 \\ \hline
\hline
IJ=11 & 142 & 144 & 145 & 146 & 232 & 23$_1$3 & 23$_2$3 & 235 &
23$_1$6 & 23$_2$6 \\ \hline
$\overline{17}$ & 4/135 & 0 & -8/135 & 2/135 & 0 & 0 & 0 & 0 & 0 & 0 \\
$\overline{26}$ & 0 & 0 & 0 & 0 & 5/108 & -1/9 & -1/36 & 5/27 & 1/27 &
-1/27 \\
$\overline{35}$ & 0 & 0 & 0 & 0 & 5/108 & 1/81 & 1/324 & 5/27 & -4/27 &
4/27 \\
$\overline{46}$ & 2/45 & 0 & 0 & -1/45 & 0 & 0 & 0 & 0 & 0 & 0 \\
$\widetilde{35}$ & 1/135 & 0 & 8/135 & 8/135 & 0 & 0 & 0 & 0 & 0 & 0 \\
$\overline{57}$ & -1/27 & -8/135 & 0 & 0 & 0 & 0 & 0 & 0 & 0 & 0 \\
$\widetilde{26}$ & 1/135 & 0 & 8/135 & -2/135 & 0 & 0 & 0 & 0 & 0 & 0 \\
$\widetilde{46}$ & 0 & 0 & 0 & 0 & 5/18 & 2/27 & 1/54 & 0 & 1/18 & -1/18
\\
66 & -2/27 & 4/135 & 0 & 0 & 0 & 0 & 0 & 0 & 0 & 0 \\
\end{tabular}}
\pagebreak
\vspace*{-1.1in}

\begin{center}
{\bf TABLE 1c Cont-}\\
\end{center}
\vspace*{0.05in}
\hspace*{-0.75in}
\small{\begin{tabular}{p{37pt}|p{37pt}p{37pt}p{37pt}p{37pt}p{37pt}p{37pt}p
{37pt}p{37pt}p{37pt}p{37pt}}
\hline
$\widetilde{17}$ & 0 & 0 & 0 & 0 & 5/27 & -4/81 & -1/81 & -5/27 & -1/27 &
1/27 \\
$\widetilde{57}$ & 0 & 0 & 0 & 0 & 0 & 20/81 & -20/81 & 0 & 0 & 0 \\
$\overline{13}$ & 0 & 0 & 0 & 0 & 0 & 2/81 & 8/81 & 0 & -1/6 & -1/6 \\
22 & 0 & 1/54 & 1/270 & 8/135 & 0 & 0 & 0 & 0 & 0 & 0 \\
$\overline{24}$ & 0 & 0 & 0 & 0 & 0 & 1/27 & 4/27 & 0 & 1/9 & 1/9 \\
$\widetilde{13}$ & 0 & 1/27 & 1/135 & -4/135 & 0 & 0 & 0 & 0 & 0 & 0 \\
$\widetilde{42}$ & 0 & 1/18 & -1/90 & 0 & 0 & 0 & 0 & 0 & 0 & 0 \\ \hline
& 322 & 323 & 324 & 325 & 326 & 342 & 344 & 345 & 346 & 413 \\ \hline
$\overline{17}$ & 5/27 & -1/5 & 0 & 1/27 & 8/135 & 16/135 & 0 & -32/135
& 8/135 & 0 \\
$\overline{26}$ & 0 & 0 & 0 & 0 & 0 & 0 & 0 & 0 & 0 & 1/5 \\
$\overline{35}$ & 0 & 0 & 0 & 0 & 0 & 0 & 0 & 0 & 0 & -1/45 \\
$\overline{46}$ & 5/18 & 3/10 & 0 & 0 & -4/45 & 8/45 & 0 & 0 & -4/45 & 0 \\
$\widetilde{35}$ & 5/108 & 1/20 & 0 & -1/27 & 32/135 & 4/135 & 0 &
32/135 & 32/135 & 0 \\
$\overline{57}$ & 4/27 & 0 & 10/27 & 0 & 0 & -4/27 & -32/135 & 0 & 0 & 0
\\ $\widetilde{26}$ & 5/108 & -9/20 & 0 & -1/27 & -8/135 & 24/135 & 0 &
32/135 & -8/135 & 0 \\
$\widetilde{46}$ & 0 & 0 & 0 & 0 & 0 & 0 & 0 & 0 & 0 & -2/15 \\
66 & 8/27 & 0 & -5/27 & 0 & 0 & -8/27 & 16/135 & 0 & 0 & 0 \\
$\widetilde{17}$ & 0 & 0 & 0 & 0 & 0 & 0 & 0 & 0 & 0 & 4/45 \\
$\widetilde{57}$ & 0 & 0 & 0 & 0 & 0 & 0 & 0 & 0 & 0 & 1/9 \\
$\overline{13}$ & 0 & 0 & 0 & 0 & 0 & 0 & 0 & 0 & 0 & 8/45 \\
22 & 0 & 0 & 2/27 & 4/27 & -10/27 & 0 & 2/27 & 2/135 & 32/135 & 0 \\
$\overline{24}$ & 0 & 0 & 0 & 0 & 0 & 0 & 0 & 0 & 0 & 4/15 \\
$\widetilde{13}$ & 0 & 0 & 4/27 & 8/27 & 5/27 & 0 & 4/27 & 4/135 &
-16/135 & 0 \\
$\widetilde{42}$ & 0 & 0 & 2/9 & -4/9 & 0 & 0 & 2/9 & -2/45 & 0 & 0 \\ \hline
& 432 & 43$_1$3 & 43$_2$3 & 435 & 43$_1$6 & 43$_2$6 \\ \hline
$\overline{17}$ & 0 & 0 & 0 & 0 & 0 & 0 \\
$\overline{26}$ & 1/27 & -4/45 & -1/45 & 4/27 & 4/135 & -4/135 \\
$\overline{35}$ & 1/27 & 4/405 & 1/405 & 4/27 & -16/135 & 16/135 \\
$\overline{46}$ & 0 & 0 & 0 & 0 & 0 & 0 \\
$\widetilde{35}$ & 0 & 0 & 0 & 0 & 0 & 0 \\
$\overline{57}$ & 0 & 0 & 0 & 0 & 0 & 0 \\
$\widetilde{26}$ & 0 & 0 & 0 & 0 & 0 & 0 \\
$\widetilde{46}$ & 2/9 & 8/135 & 2/135 & 0 & 2/45 & -2/45  \\
66 & 0 & 0 & 0 & 0 & 0 & 0 \\
$\widetilde{17}$ & 4/27 & -16/405 & -4/405 & -4/27 & -4/135 & 4/135 \\
$\widetilde{57}$ & 0 & 16/81 & -16/81 & 0 & 0 & 0 \\
$\overline{13}$ & 0 & 8/405 & 32/405 & 0 & -2/15 & -2/15 \\
22 & 0 & 0 & 0 & 0 & 0 & 0 \\
$\overline{24}$ & 0 & 4/135 & 16/135 & 0 & 4/45 & 4/45 \\
$\widetilde{13}$ & 0 & 0 & 0 & 0 & 0 & 0 \\
$\widetilde{42}$ & 0 & 0 & 0 & 0 & 0 & 0 \\ \hline
\end{tabular}}

\pagebreak
\vspace*{-1.1in}
\begin{center}
{\bf TABLE 1c Cont-}\\
\vspace*{0.05in}
\hspace*{-0.5in}
\small{\begin{tabular}{p{35pt}|p{37pt}p{35pt}p{35pt}p{35pt}p{35pt}p{35pt}p
{35pt}p{35pt}p{35pt}p{35pt}p{35pt}}
\hline
IJ=10 & 143 & 232 & 234 & 235 & 236 & 323 & 326 & 343 & 414 & 432 \\
\hline
$\overline{13}$ & 2/45 & 0 & 0 & 0 & 0 & 8/45 & 3/5 & 8/45 & 0 & 0 \\
$\overline{24}$ & 1/15 & 0 & 0 & 0 & 0 & 4/15 & -2/5 & 4/15 & 0 & 0 \\
$\widetilde{13}$ & 0 & 0 & 1/27 & -5/27 & -5/27 & 0 & 0 & 0 & 4/15 & 0 \\
$\widetilde{42}$ & 0 & 0 & 1/18 & 5/18 & 0 & 0 & 0 & 0 & 2/5 & 0 \\
$\overline{57}$ & 0 & -5/27 & -8/27 & 0 & 0 & 0 & 0 & 0 & 2/15 & -4/27 \\
$\widetilde{57}$ & -4/45 & 0 & 0 & 0 & 0 & 5/9 & 0 & -16/45 & 0 & 0 \\
66 & 0 & -10/27 & 4/27 & 0 & 0 & 0 & 0 & 0 & -1/15 & -8/27 \\
22 & 0 & 0 & 1/54 & -5/54 & 10/27 & 0 & 0 & 0 & 2/15 & 0 \\ \hline
& 434 & 435 & 436 \\ \hline
$\overline{13}$ & 0 & 0 & 0 \\
$\overline{24}$ & 0 & 0 & 0 \\
$\widetilde{13}$ & 4/135 & -4/27 & -4/27 \\
$\widetilde{42}$ & 2/45 & 2/9 & 0 \\
$\overline{57}$ & -32/135 & 0 & 0 \\
$\widetilde{57}$ & 0 & 0 & 0 \\
66 & 16/135 & 0 & 0 \\
22 & 2/135 & -2/27 & 8/27 \\ \hline
\hline
IJ=03 & 233 & 433 \\ \hline
66 & -5/9 & -4/9 \\ \hline
\hline
IJ=02 & 143 & 146 & 233 & 236 & 323 & 343 & 346 & 433 & 436 \\\hline
$\overline{17}$ & 4/75 & -2/25 & 0 & 0 & 1/3 & 16/75 & -8/25 & 0 & 0 \\
$\overline{26}$ & 0 & 0 & 2/9 & 1/3 & 0 & 0 & 0 & 8/45 & 4/15 \\
$\widetilde{26}$ & 8/225 & 3/25 & 0 & 0 & 2/9 & 32/225 & 12/25 & 0 & 0 \\
66 & -1/9 & 0 & 0 & 0 & 4/9 & -4/9 & 0 & 0 & 0 \\
$\widetilde{17}$ & 0 & 0 & 1/3 & -2/9 & 0 & 0 & 0 & 4/15 & -8/45 \\ \hline
\hline
IJ=01 & 146 & 23$_1$3 & 23$_2$3 & 23$_1$6 & 23$_2$6 & 237 & 323 &
326 & 346 & 413 \\ \hline
$\overline{13}$ & 4/45 & 0 & 0 & 0 & 0 & 0 & 0 & -5/9 & 16/45 & 0 \\
$\overline{17}$ & 0 & -4/27 & -1/27 & -1/9 & 1/9 & 0 & 0 & 0 & 0 & 4/15 \\
$\overline{26}$ & -1/15 & 0 & 0 & 0 & 0 & 0 & -2/5 & -4/15 & -4/15 & 0 \\
$\widetilde{13}$ & 0 & -1/54 & -2/27 & -1/18 & -1/18 & 5/18 & 0 & 0 & 0 &
-2/15 \\
$\widetilde{26}$ & 0 & -8/81 & -2/81 & 1/6 & -1/6 & 0 & 0 & 0 & 0 & 8/45 \\
$\widetilde{17}$ & 2/45 & 0 & 0 & 0 & 0 & 0 & -3/5 & 8/45 & 8/45 & 0 \\
22 & 0 & -1/648 & -1/162 & -1/6 & -1/6 & -5/24 & 0 & 0 & 0 & -1/90 \\
44 & 0 & 1/24 & 1/6 & -1/18 & -1/18 & 5/72 & 0 & 0 & 0 & 3/10 \\
66 & 0 & 20/81 & -20/81 & 0 & 0 & 0 & 0 & 0 & 0 & 1/9 \\ \hline
\end{tabular}}
\end{center}

\pagebreak
\vspace*{-1.1in}
\begin{center}
{\bf TABLE 1c Cont-}\\
\vspace*{-0.05in}
\end{center}
\small{\begin{tabular}{p{35pt}|p{37pt}p{35pt}p{35pt}p{35pt}p{35pt}p{35pt}p
{35pt}p{35pt}p{35pt}p{35pt}p{35pt}}
\hline
& 43$_1$3 & 43$_2$3 & 43$_1$6 & 43$_2$6 & 437 \\ \hline
$\overline{13}$ & 0 & 0 & 0 & 0 & 0 \\
$\overline{17}$ & -16/135 & -4/135 & -4/45 & 4/45 & 0 \\
$\overline{26}$ & 0 & 0 & 0 & 0 & 0 \\
$\widetilde{13}$ & -2/135 & -8/135 & -2/45 & -2/45 & 2/9 \\
$\widetilde{26}$ & -32/405 & -8/405 & 2/15 & -2/15 & 0 \\
$\widetilde{17}$ & 0 & 0 & 0 & 0 & 0 \\
22 & -1/810 & -2/405 & -2/15 & -2/15 & -1/6 \\
44 & 1/30 & 2/15 & -2/45 & -2/45 & 1/18 \\
66 & 16/81 & -16/81 & 0 & 0 & 0 \\ \hline
\hline
IJ=00 & 143 & 147 & 236 & 323 & 326 & 343 & 347 & 436 \\ \hline
$\overline{13}$ & 0 & 0 & 5/9 & 0 & 0 & 0 & 0 & 4/9 \\
22 & -1/360 & 3/40 & 0 & -1/90 & 3/5 & -1/90 & 3/10 & 0 \\
$\widetilde{13}$ & -1/30 & -1/10 & 0 & -2/15 & 1/5 & -2/15 & -2/5 & 0 \\
44 & 3/40 & -1/40 & 0 & 3/10 & 1/5 & 3/10 & -1/10 & 0 \\
66 & -4/45 & 0 & 0 & 5/9 & 0 & -16/45 & 0 & 0 \\ \hline
\end{tabular}}

\pagebreak
\vspace*{-1.1in}
\begin{center}
{\bf TABLE 1d}\\
\vspace*{0.05in}
\hspace*{-0.75in}
\small{\begin{tabular}{p{35pt}|p{37pt}p{35pt}p{35pt}p{35pt}p{35pt}p{35pt}p
{35pt}p{35pt}p{35pt}p{35pt}p{35pt}}
\hline
IJ=$\frac{3}{2}$3 & 144 & 233 & 322 & 344 & 411 & 433 \\ \hline
$\overline{58}$ & 0 & -1/2 & 0 & 0 & 1/10 & -2/5 \\
$\overline{67}$ & 0 & 1/18 & 0 & 0 & 9/10 & 2/45 \\
$\widetilde{67}$ & 1/10 & 0 & 1/2 & 2/5 & 0 & 0 \\
$\widetilde{58}$ & -1/10 & 0 & 1/2 & -2/5 & 0 & 0 \\ \hline
\hline
IJ=$\frac{3}{2}$2 & 143 & 232 & 233 & 234 & 321 & 322 & 323 & 343 &
412 & 432 \\ \hline
$\overline{27}$ & 0 & 1/18 & 5/18 & 0 & 0 & 0 & 0 & 0 & 2/5 & 2/45 \\
$\overline{36}$ & 0 & 1/18 & -5/18 & 0 & 0 & 0 & 0 & 0 & 2/5 & 2/45 \\
$\overline{58}$ & -1/10 & 0 & 0 & 0 & 1/10 & 0 & 2/5 & -2/5 & 0 & 0 \\
$\widetilde{36}$ & -2/45 & 0 & 0 & 0 & 0 & -1/2 & -5/18 & -8/45 & 0 & 0 \\
$\overline{67}$ & 1/90 & 0 & 0 & 0 & 9/10 & 0 & -2/45 & 2/45 & 0 & 0 \\
$\widetilde{27}$ & 2/45 & 0 & 0 & 0 & 0 & -1/2 & 5/18 & 8/45 & 0 & 0 \\
$\widetilde{67}$ & 0 & -2/9 & 0 & 5/18 & 0 & 0 & 0 & 0 & 1/10 & -8/45 \\
$\widetilde{58}$ & 0 & -2/9 & 0 & -5/18 & 0 & 0 & 0 & 0 & 1/10 & -8/45 \\
\hline
& 433 & 434 \\ \hline
$\overline{27}$ & 2/9 & 0 \\
$\overline{36}$ & -2/9 & 0 \\
$\overline{58}$ & 0 & 0 \\
$\widetilde{36}$ & 0 & 0 \\
$\overline{67}$ & 0 & 0 \\
$\widetilde{27}$ & 0 & 0 \\
$\widetilde{67}$ & 0 & 2/9 \\
$\widetilde{58}$ & 0 & -2/9 \\ \hline
\hline
IJ=$\frac{3}{2}$1 & 142 & 144 & 231 & 232 & 23$_1$3 & 23$_2$3 & 322
& 323 & 324 & 342 \\ \hline
$\overline{27}$ & 2/45 & 0 & 0 & 0 & 0 & 0 & 5/18 & -1/2 & 0 & 8/45 \\
$\overline{36}$ & 2/45 & 0 & 0 & 0 & 0 & 0 & 5/18 & 1/2 & 0 & 8/45 \\
$\overline{58}$ & 0 & 0 & -1/18 & 0 & 2/9 & -2/9 & 0 & 0 & 0 & 0 \\
$\widetilde{36}$ & 0 & 0 & 0 & 5/18 & 10/81 & 5/162 & 0 & 0 & 0 & 0 & \\
$\overline{67}$ & 0 & 0 & -1/2 & 0 & -2/81 & 2/81 & 0 & 0 & 0 & 0 \\
$\widetilde{27}$ & 0 & 0 & 0 & 5/18 & -10/81 & -5/162 & 0 & 0 & 0 & 0 \\
$\widetilde{67}$ & -1/18 & 2/45 & 0 & 0 & 0 & 0 & 2/9 & 0 & -5/18 & -2/9 \\
$\widetilde{58}$ & -1/18 & -2/45 & 0 & 0 & 0 & 0 & 2/9 & 0 & 5/18 & -2/9 \\
$\overline{23}$ & 0 & 0 & 0 & 0 & 5/81 & 20/81 & 0 & 0 & 0 & 0 \\
$\widetilde{23}$ & 0 & 1/9 & 0 & 0 & 0 & 0 & 0 & 0 & 4/9 & 0 \\ \hline
\end{tabular}}
\end{center}

\pagebreak
\vspace*{-1.1in}
\begin{center}
{\bf TABLE 1d Cont-}\\
\vspace*{0.05in}
\hspace*{-0.75in}
\small{\begin{tabular}{p{35pt}|p{37pt}p{35pt}p{35pt}p{35pt}p{35pt}p{35pt}p
{35pt}p{35pt}p{35pt}p{35pt}p{35pt}}
\hline
& 344 & 413 & 431 & 432 & 43$_1$3 & 43$_2$3 \\ \hline
$\overline{27}$ & 0 & 0 & 0 & 0 & 0 & 0 \\
$\overline{36}$ & 0 & 0 & 0 & 0 & 0 & 0 \\
$\overline{58}$ & 0 & 1/10 & -2/45 & 0 & 8/45 & -8/45 \\
$\widetilde{36}$ & 0 & -2/9 & 0 & 2/9 & 8/81 & 2/81 \\
$\overline{67}$ & 0 & -1/90 & -2/5 & 0 & -8/405 & 8/405 \\
$\widetilde{27}$ & 0 & 2/9 & 0 & 2/9 & -8/81 & -2/81 \\
$\widetilde{67}$ & 8/45 & 0 & 0 & 0 & 0 & 0 \\
$\widetilde{58}$ & -8/45 & 0 & 0 & 0 & 0 & 0 \\
$\overline{23}$ & 0 & 4/9 & 0 & 0 & 4/81 & 16/81 \\
$\widetilde{23}$ & 4/9 & 0 & 0 & 0 & 0 & 0 \\ \hline
\hline
IJ=$\frac{3}{2}$0 & 141 & 143 & 232 & 234 & 323 & 341 & 343 & 414 &
432 & 434 \\ \hline
$\overline{58}$ & -1/50 & -2/25 & 0 & 0 & 1/2 & -2/25 & -8/25 & 0 & 0 & 0
\\
$\overline{67}$ & -9/50 & 2/225 & 0 & 0 & -1/18 & -18/25 & 8/225 & 0 & 0
& 0 \\
$\widetilde{67}$ & 0 & 0 & -5/18 & 2/9 & 0 & 0 & 0 & -1/10 & -2/9 & 8/45 \\
$\widetilde{58}$ & 0 & 0 & -5/18 & -2/9 & 0 & 0 & 0 & 1/10 & -2/9 & -8/45 \\
$\overline{23}$ & 0 & 1/9 & 0 & 0 & 4/9 & 0 & 4/9 & 0 & 0 & 0 \\
$\widetilde{23}$ & 0 & 0 & 0 & 1/9 & 0 & 0 & 0 & 4/5 & 0 & 4/45 \\ \hline
\hline
IJ=$\frac{1}{2}$3 & 233 & 322 & 433 \\ \hline
$\overline{67}$ & 0 & 1 & 0 \\
$\widetilde{67}$ & -5/9 & 0 & -4/9 \\ \hline
\hline
IJ=$\frac{1}{2}$2 & 143 & 146 & 232 & 233 & 235 & 236 & 322 & 323 &
325 & 343 \\ \hline
$\overline{18}$ & 0 & 0 & 1/72 & 1/8 & -5/36 & -2/9 & 0 & 0 & 0 & 0 \\
$\overline{27}$ & 49/900 & 1/25 & 0 & 0 & 0 & 0 & -1/16 & 49/144 & -1/8 &
49/225 \\
$\overline{36}$ & -1/225 & -1/25 & 0 & 0 & 0 & 0 & -1/4 & -1/36 & -1/2 &
-4/225 \\
$\overline{47}$ & 0 & 0 & 1/16 & -1/16 & -5/72 & 1/9 & 0 & 0 & 0 & 0 \\
$\widetilde{36}$ & 0 & 0 & 1/36 & -1/36 & 5/18 & -1/9 & 0 & 0 & 0 & 0 \\
$\overline{67}$ & 0 & 0 & -4/9 & 0 & 0 & 0 & 0 & 0 & 0 & 0 \\
$\widetilde{27}$ & 0 & 0 & 1/144 & 49/144 & 5/72 & 1/9 & 0 & 0 & 0 & 0 \\
$\widetilde{47}$ & -1/100 & 1/25 & 0 & 0 & 0 & 0 & -9/16 & -1/16 & 1/8 &
-1/25 \\
$\widetilde{67}$ & -1/9 & 0 & 0 & 0 & 0 & 0 & 0 & 4/9 & 0 & -4/9 \\
$\widetilde{18}$ & 1/50 & -2/25 & 0 & 0 & 0 & 0 & -1/8 & 1/8 & 1/4 & 2/25
\\ \hline
& 346 & 412 & 432 & 433 & 435 & 436 \\ \hline
$\overline{18}$ & 0 & 1/10 & 1/90 & 1/10 & -1/9 & -8/45 \\
$\overline{27}$ & 4/25 & 0 & 0 & 0 & 0 & 0 \\
$\overline{36}$ & -4/25 & 0 & 0 & 0 & 0 & 0 \\
$\overline{47}$ & 0 & 9/20 & 1/20 & -1/20 & -1/18 & 4/45 \\
$\widetilde{36}$ & 0 & 1/5 & 1/45 & -1/45 & 2/9 & -4/45 \\
$\overline{67}$ & 0 & 1/5 & -16/45 & 0 & 0 & 0 \\
$\widetilde{27}$ & 0 & 1/20 & 1/180 & 49/180 & 1/18 & 4/45 \\
$\widetilde{47}$ & 4/25 & 0 & 0 & 0 & 0 & 0 \\
$\widetilde{67}$ & 0 & 0 & 0 & 0 & 0 & 0 \\
$\widetilde{18}$ & -8/25 & 0 & 0 & 0 & 0 & 0 \\ \hline
\end{tabular}}
\end{center}

\pagebreak
\vspace*{-1.1in}
\begin{center}
{\bf TABLE 1d Cont-}\\
\vspace*{0.05in}
\hspace*{-0.75in}
\small{\begin{tabular}{p{30pt}|p{32pt}p{30pt}p{30pt}p{30pt}p{37pt}p{42pt}p
{37pt}p{40pt}p{42pt}p{30pt}p{30pt}}
\hline
IJ=$\frac{1}{2}$0 & 143 & 232 & 235 & 236 & 323 & 326 &
343 & 432 & 435 & 436 \\
\hline
$\overline{23}$ & 0 & 0 & -5/18 & 5/18 & 0 & 0 & 0 & 0
& -2/9 & 2/9 \\
$\widetilde{23}$ & -1/90 & 0 & 0 & 0 & -2/45 & 9/10 &
-2/45 & 0 & 0 & 0 \\
$\overline{34}$ & 1/10 & 0 & 0 & 0 & 2/5 & 1/10 & 2/5 &
0 & 0 & 0 \\
$\widetilde{34}$ & 0 & 0 & 5/18 & 5/18 & 0 & 0 & 0 & 0
& 2/9 & 2/9 \\
$\overline{67}$ & 0 & -5/9 & 0 & 0 & 0 & 0 & 0 & -4/9 &
0 & 0 & \\
$\widetilde{67}$ & -4/45 & 0 & 0 & 0 & 5/9 & 0 & -16/45
& 0 & 0 & 0 \\
\hline\hline
IJ=$\frac{1}{2}$1 & 142 & 145 & 146 & 232 & 23$_1$3 &
23$_2$3 & 235 & 23$_1$6 & 23$_2$6 & 322 \\
\hline
$\overline{18}$ & 1/90 & -2/45 & 2/45 & 0 & 0 & 0 & 0 &
0 & 0 & 5/72 \\
$\overline{27}$ & 0 & 0 & 0 & 5/144 & -49/324 &
-49/1296 & 5/72 & 1/18 & -1/18 & 0 \\
$\overline{36}$ & 0 & 0 & 0 & 5/36 & 1/81 & 1/324 &
5/18 & -1/18 & 1/18 & 0 \\
$\overline{47}$ & 1/20 & -1/45 & -1/45 & 0 & 0 & 0 & 0
& 0 & 0 & 5/16 \\
$\widetilde{36}$ & 1/45 & 4/45 & 1/45 & 0 & 0 & 0 & 0 &
0 & 0 & 5/36 \\
$\overline{67}$ & -1/9 & 0 & 0 & 0 & 0 & 0 & 0 & 0 & 0
& 4/9 \\
$\widetilde{27}$ & 1/180 & 1/45 & -1/45 & 0 & 0 & 0 & 0
& 0 & 0 & 5/144 \\
$\widetilde{47}$ & 0 & 0 & 0 & 5/16 & 1/36 & 1/144 &
-5/72 & 1/18 & -1/18 & 0 \\
$\widetilde{67}$ & 0 & 0 & 0 & 0 & 20/81 & -20/81 & 0 &
0 & 0 & 0 \\
$\widetilde{18}$ & 0 & 0 & 0 & 5/72 & -1/18 & -1/72 &
-5/36 & -1/9 & 1/9 & 0 \\
$\overline{23}$ & 0 & 1/90 & 2/45 & 0 & 0 & 0 & 0 & 0 &
0 & 0 \\
$\widetilde{23}$ & 0 & 0 & 0 & 0 & -1/162 & -2/81 & 0 &
-1/4 & -1/4 & 0 \\
$\overline{34}$ & 0 & 0 & 0 & 0 & 1/18 & 2/9 & 0 &
-1/36 & -1/36 & 0 \\
$\widetilde{34}$ & 0 & -1/90 & 2/45 & 0 & 0 & 0 & 0 & 0
& 0 & 0 \\
\hline
& 323 & 325 & 326 & 342 & 345 & 346 & 413 & 432 &
43$_1$3 & 43$_2$3 \\
\hline
$\overline{18}$ & -9/40 & 1/36 & 2/45 & -8/45 & 8/45 &
0 & 0 & 0 & 0 & 0 \\
$\overline{27}$ & 0 & 0 & 0 & 0 & 0 & 49/180 & 1/36 &
-49/405 & -49/1620 & 1/18 \\
$\overline{36}$ & 0 & 0 & 0 & 0 & 0 & -1/45 & 1/9 &
4/405 & 1/405 & 2/9 \\
$\overline{47}$ & 9/80 & 1/72 & 1/5 & -4/45 & -4/45 & 0
& 0 & 0 & 0 & 0 \\
$\widetilde{36}$ & 1/20 & -1/18 & 4/45 & 16/45 & 4/45 &
0 & 0 & 0 & 0 & 0 \\
$\overline{67}$ & 0 & 0 & -4/9 & 0 & 0 & 0 & 0 & 0 & 0
& 0 \\
$\widetilde{27}$ & -49/80 & -1/72 & 1/45 & 4/45 & -4/45
& 0 & 0 & 0 & 0 & 0 \\
$\widetilde{47}$ & 0 & 0 & 0 & 0 & 0 & -1/20 & 1/4 &
1/45 & 1/180 & -1/18 \\
$\widetilde{67}$ & 0 & 0 & 0 & 0 & 0 & 1/9 & 0 & 16/81
& -16/81 & 0 \\
$\widetilde{18}$ & 0 & 0 & 0 & 0 & 0 & 1/10 & 1/18 &
-2/45 & -1/90 & -1/9 \\
$\overline{23}$ & 0 & 4/9 & 0 & 2/45 & 8/45 & 0 & 0 & 0
& 0 & 0 \\
$\widetilde{23}$ & 0 & 0 & 0 & 0 & 0 & -2/45 & 0 &
-2/405 & -8/405 & 0 \\
$\overline{34}$ & 0 & 0 & 0 & 0 & 0 & 2/5 & 0 & 2/45 &
8/45 & 0 \\
$\widetilde{34}$ & 0 & -4/9 & 0 & -2/45 & 8/45 & 0 & 0
& 0 & 0 & 0 \\
\hline
\end{tabular}}
\end{center}

\pagebreak
\vspace*{-1.1in}
\begin{center}
{\bf TABLE 1d Cont-}\\
\vspace*{0.05in}
\small{\begin{tabular}{p{30pt}|p{32pt}p{30pt}p{30pt}p{30pt}p{37pt}p{42pt}p
{37pt}p{40pt}p{42pt}p{30pt}p{30pt}p{30pt}p{30pt}}
\hline
& 435 & 43$_1$6 & 43$_2$6 \\ \hline
$\overline{18}$ & 0 & 0 & 8/45 \\
$\overline{27}$ & 2/45 & -2/45 & 0 \\
$\overline{36}$ & -2/45 & 2/45 & 0 \\
$\overline{47}$ & 0 & 0 & -4/45 \\
$\widetilde{36}$ & 0 & 0 & 4/45 \\
$\overline{67}$ & 0 & 0 & 0 \\
$\widetilde{27}$ & 0 & 0 & -4/45 \\
$\widetilde{47}$ & 2/45 & -2/45 & 0 \\
$\widetilde{67}$ & 0 & 0 & 0 \\
$\widetilde{18}$ & -4/45 & 4/45 & 0 \\
$\overline{23}$ & 0 & 0 & -5/18 \\
$\widetilde{23}$ & -1/5 & -1/5 & 0 \\
$\overline{34}$ & -1/45 & -1/45 & 0 \\
$\widetilde{34}$ & 0 & 0 & -5/18 \\ \hline
\end{tabular}}
\end{center}

\pagebreak
\vspace*{-1.1in}
\begin{center}
{\bf TABLE 1e}\\
\vspace*{-0.05in}
\hspace*{-0.25in}
\end{center}
\small{\begin{tabular}{p{30pt}|p{30pt}p{37pt}p{37pt}p{30pt}p{33pt}p{33pt}p
{30pt}p{30pt}p{32pt}p{30pt}p{30pt}p{30pt}p{30pt}}
\hline
IJ=13 & 233 & 322 & 411 & 433 \\ \hline
$\overline{68}$ & -1/3 & 0 & 2/5 & -4/15 \\
77 & 2/9 & 0 & 3/5 & 8/45 \\
$\widetilde{68}$ & 0 & 1 & 0 & 0 \\ \hline
\hline
IJ=12 & 143 & 232 & 233 & 321 & 322 & 323 & 343 & 412 & 432 & 433 \\
\hline
$\overline{28}$ & 0 & 1/36 & 5/12 & 0 & 0 & 0 & 0 & 1/5 & 1/45 & 1/3 \\
$\overline{37}$ & 0 & 1/12 & -5/36 & 0 & 0 & 0 & 0 & 3/5 & 1/15 & -1/9 \\
$\overline{68}$ & -1/15 & 0 & 0 & 2/5 & 0 & 4/15 & -4/15 & 0 & 0 & 0 \\
$\widetilde{37}$ & -1/45 & 0 & 0 & 0 & -3/4 & -5/36 & -4/45 & 0 & 0 & 0 \\
77 & 2/45 & 0 & 0 & 3/5 & 0 & -8/45 & 8/45 & 0 & 0 & 0 \\
$\widetilde{28}$ & 1/15 & 0 & 0 & 0 & -1/4 & 5/12 & 4/15 & 0 & 0 & 0 \\
$\widetilde{68}$ & 0 & -4/9 & 0 & 0 & 0 & 0 & 0 & 1/5 & -16/45 & 0 \\ \hline
\hline
IJ=11 & 142 & 231 & 232 & 23$_1$3 & 23$_2$3 & 322 & 323 & 342 &
413 & 431 \\ \hline
$\overline{28}$ & 1/45 & 0 & 0 & 0 & 0 & 5/36 & -3/4 & 4/45 & 0 & 0 \\
$\overline{37}$ & 1/15 & 0 & 0 & 0 & 0 & 5/12 & 1/4 & 4/15 & 0 & 0 \\
$\overline{68}$ & 0 & -2/9 & 0 & 4/27 & -4/27 & 0 & 0 & 0 & 1/15 & -8/45 \\
$\widetilde{37}$ & 0 & 0 & 5/12 & 5/81 & 5/324 & 0 & 0 & 0 & -1/9 & 0 \\
$\widetilde{28}$ & 0 & 0 & 5/36 & -5/27 & -5/108 & 0 & 0 & 0 & 1/3 & 0 \\
$\widetilde{68}$ & -1/9 & 0 & 0 & 0 & 0 & 4/9 & 0 & -4/9 & 0 & 0 \\
77 & 0 & -1/3 & 0 & -8/81 & 8/81 & 0 & 0 & 0 & -2/45 & -4/15 \\
33 & 0 & 0 & 0 & 5/81 & 20/81 & 0 & 0 & 0 & 4/9 & 0 \\ \hline
& 432 & 43$_1$3 & 43$_2$3 \\ \hline
$\overline{28}$ & 0 & 0 & 0 \\
$\overline{37}$ & 0 & 0 & 0 \\
$\overline{68}$ & 0 & 16/135 & -16/135 \\
$\widetilde{37}$ & 1/3 & 4/81 & 1/81 \\
$\widetilde{28}$ & 1/9 & -4/27 & -1/27 \\
$\widetilde{68}$ & 0 & 0 & 0 \\
77 & 0 & -32/405 & 32/405 \\
33 & 0 & 4/81 & 16/81 \\ \hline
\hline
IJ=10 & 141 & 143 & 232 & 323 & 341 & 343 & 432 \\ \hline
$\overline{68}$ & -2/25 & -4/75 & 0 & 1/3 & -8/25 & -16/75 & 0 \\
77 & -3/25 & 8/225 & 0 & -2/9 & -12/25 & 32/225 & 0 \\
$\widetilde{68}$ & 0 & 0 & -5/9 & 0 & 0 & 0 & -4/9 \\
33 & 0 & 1/9 & 0 & 4/9 & 0 & 4/9 & 0 \\ \hline
\hline
IJ=03 & 322 \\ \hline
77 & 1 \\ \hline
\hline
IJ=02 & 232 & 235 & 322 & 325 & 412 & 432 & 435 \\ \hline
$\overline{37}$ & 0 & 0 & -1/2 & -1/2 & 0 & 0 & 0 \\
$\overline{48}$ & 1/18 & -5/18 & 0 & 0 & 2/5 & 2/45 & -2/9 \\
$\widetilde{37}$ & 1/18 & 5/18 & 0 & 0 & 2/5 & 2/45 & 2/9 \\
77 & -4/9 & 0 & 0 & 0 & 1/5 & -16/45 & 0 \\
$\widetilde{48}$ & 0 & 0 & -1/2 & 1/2 & 0 & 0 & 0 \\ \hline
\end{tabular}}

\pagebreak
\vspace*{-1.1in}
\begin{center}
{\bf TABLE 1e Cont-}\\
\vspace*{-0.05in}
\end{center}
\small{\begin{tabular}{p{30pt}|p{30pt}p{30pt}p{30pt}p{30pt}p{30pt}p{30pt}p
{30pt}p{35pt}p{30pt}p{30pt}p{30pt}p{30pt}p{30pt}}
\hline
IJ=01 & 142 & 145 & 232 & 235 & 322 & 325 & 342 & 345 & 432 & 435 \\
\hline
$\overline{37}$ & 0 & 0 & 5/18 & 5/18 & 0 & 0 & 0 & 0 & 2/9 & 2/9 \\
$\overline{48}$ & 2/45 & -4/45 & 0 & 0 & 5/18 & 1/18 & 8/45 & -16/45 & 0
& 0 \\
$\widetilde{37}$ & 2/45 & 4/45 & 0 & 0 & 5/18 & -1/18 & 8/45 & 16/45 & 0
& 0 \\
77 & -1/9 & 0 & 0 & 0 & 4/9 & 0 & -4/9 & 0 & 0 & 0 \\
$\widetilde{48}$ & 0 & 0 & 5/18 & -5/18 & 0 & 0 & 0 & 0 & 2/9 & -2/9 \\
33 & 0 & 1/45 & 0 & 0 & 0 & 8/9 & 0 & 4/45 & 0 & 0 \\ \hline
\hline
IJ=00 & 232 & 235 & 432 & 435 \\ \hline
77 & -5/9 & 0 & -4/9 & 0 \\
33 & 0 & -5/9 & 0 & -4/9 \\ \hline
\end{tabular}}

\pagebreak
\vspace*{-1.1in}
\begin{center}
{\bf TABLE 1f}\\
\vspace*{0.05in}
\small{\begin{tabular}{p{30pt}|p{30pt}p{30pt}p{30pt}p{30pt}p{35pt}p{30pt}p
{30pt}p{35pt}}
\hline
IJ=$\frac{1}{2}$3 & 322 & 411 \\ \hline
$\overline{78}$ & 0 & 1 \\

$\widetilde{78}$ & 1 & 0 \\ \hline
\hline
IJ=$\frac{1}{2}$2 & 232 & 321 & 322 & 412 & 432 \\ \hline
$\overline{38}$ & 1/9 & 0 & 0 & 4/5 & 4/45 \\
$\overline{78}$ & 0 & 1 & 0 & 0 & 0 \\
$\widetilde{38}$ & 0 & 0 & -1 & 0 & 0 \\
$\widetilde{78}$ & -4/9 & 0 & 0 & 1/5 & -16/45 \\ \hline
\hline
IJ=$\frac{1}{2}$1 & 142 & 231 & 232 & 322 & 342 & 431 & 432 \\ \hline
$\overline{38}$ & 4/45 & 0 & 0 & 5/9 & 16/45 & 0 & 0 \\
$\overline{78}$ & 0 & -5/9 & 0 & 0 & 0 & -4/9 & 0 \\
$\widetilde{38}$ & 0 & 0 & 5/9 & 0 & 0 & 0 & 4/9 \\
$\widetilde{78}$ & -1/9 & 0 & 0 & 4/9 & -4/9 & 0 & 0 \\ \hline
\hline
IJ=$\frac{1}{2}$0 & 141 & 232 & 341 & 342 \\ \hline
$\overline{78}$ & -1/5 & 0 & -4/5 & 0 \\
$\widetilde{78}$ & 0 & -5/9 & 0 & -4/9 \\ \hline
\end{tabular}}
\end{center}

\pagebreak
\vspace*{-1.1in}
\begin{center}
{\bf TABLE 1g}\\
\vspace*{0.15in}
\small{\begin{tabular}{p{30pt}|p{30pt}p{30pt}p{30pt}p{30pt}p{35pt}p{30pt}p
{30pt}p{35pt}}
\hline
IJ=03 & 411 \\ \hline
88 & 1 \\ \hline
\hline
IJ=02 & 321 \\ \hline
88 & 1 \\ \hline
\hline
IJ=01 & 231 & 431 \\ \hline
88 & -5/9 & -4/9 \\ \hline
\hline
IJ=00 & 141 & 341 \\ \hline
88 & -1/5 & -4/5 \\ \hline
\end{tabular}}
\end{center}

\pagebreak
\vspace*{-1.1in}
\begin{center}
{\bf TABLE 2} \\
\end{center}
\begin{tabular*}{5.86in}[c]{|p{45pt}|p{30pt}|p{25pt}|p{50pt}|p{30pt}|p{25pt}|
p{55pt}|p{30pt}|p{25pt}|}
\hline
{$\nu_1 \nu_2$} & {$\nu$} & {$\epsilon_2$} & {$\nu_1 \nu_2$} & {$\nu$} &
{$\epsilon_2$} & {$\nu_1 \nu_2$} & {$\nu$} & {$\epsilon_2$} \\
\hline
$[$4$]$ $[$11$]$ & $[$51$]$ & 1 & $[$22$]$ $[$11$]$ & $[$33$]$ & 1 &
$[$211$]$ $[$11$]$ & $[$2211$]$ & 1 \\
& $[$411$]$ & 1 & & $[$321$]$ & -1 & & $[$21$^4$$]$ & 1 \\
$[$31$]$ $[$11$]$ & $[$42$]$ & 1 & & $[$2211$]$ & -1 & $[$1$^4$$]$
$[$11$]$ & $[$2211$]$ & 1 \\
& $[$321$]$ & 1 & $[$211$]$$[$11$]$ & $[$321$]$ & 1 & & $[$21$^4$$]$
& -1 \\
& $[$31$^3$$]$ & -1 & & $[$222$]$ & -1 & & $[$1$^6$$]$ & 1 \\
& $[$411$]$ & -1 & & $[$31$^3$$]$ & 1 & & & \\ \hline

\end{tabular*}
\end{document}